\documentclass[pdflatex,sn-nature]{sn-jnl}

\usepackage{graphicx}%
\usepackage{multirow}%
\usepackage{amsmath,amssymb,amsfonts}%
\usepackage{amsthm}%
\usepackage{mathrsfs}%
\usepackage[title]{appendix}%
\usepackage{xcolor}%
\usepackage{textcomp}%
\usepackage{manyfoot}%
\usepackage{booktabs}%
\usepackage{algorithm}%
\usepackage{algorithmicx}%
\usepackage{algpseudocode}%
\usepackage{listings}%

\theoremstyle{thmstyleone}%
\theoremstyle{thmstyletwo}%

\theoremstyle{thmstylethree}%

\begin{document}

\title[Article Title]{Latitudinal chemical and cloud variations in the atmosphere of a brown dwarf}


\author*[1,2]{\fnm{Benjamin} \sur{Charnay}}\email{benjamin.charnay@obspm.fr}

\author[3]{\fnm{Sam} \sur{de Regt}}

\author[4,5,6]{\fnm{Matthieu} \sur{Ravet}}

\author[1]{\fnm{Lucas} \sur{Teinturier}}

\author[1]{\fnm{Flavien} \sur{Kiefer}}

\author[6]{\fnm{Gaël} \sur{Chauvin}}

\author[7]{\fnm{Allan} \sur{Denis}}

\author[5]{\fnm{Mickaël} \sur{Bonnefoy}}

\author[1]{\fnm{Paulina} \sur{Palma-Bifani}}

\author[1]{\fnm{Alice} \sur{Radcliffe}}

\author[7]{\fnm{Arthur} \sur{Vigan}}


\affil[1]{\orgdiv{LIRA}, \orgname{Observatoire de Paris, Université PSL, CNRS, Sorbonne Université, Université Paris Cité}, \orgaddress{\street{5 place Jules Janssen}, \city{Meudon}, \postcode{92195}, \country{France}}}

\affil[2]{\orgdiv{Laboratoire d’astrophysique de Bordeaux}, \orgname{Univ. Bordeaux, CNRS}, \orgaddress{\street{B18N, allée Geoffroy Saint-Hilaire}, \city{Pessac}, \postcode{33615}, \country{France}}}

\affil[3]{\orgdiv{Leiden Observatory}, \orgname{Leiden University}, \orgaddress{\street{PO Box 9513}, \city{Leiden}, \postcode{2300 RA}, \country{The Netherlands}}}

\affil[4]{\orgdiv{Laboratoire J.-L. Lagrange}, \orgname{Université Côte d’Azur, Observatoire dela Côte d’Azur, CNRS}, \orgaddress{\city{Nice}, \postcode{06304}, \country{France}}}

\affil[5]{\orgdiv{IPAG}, \orgname{Université   Grenoble-Alpes, CNRS}, \orgaddress{\city{Grenoble}, \postcode{F-38000}, \country{France}}}

\affil[6]{\orgdiv{Max-Planck-Institut für Astronomie}, \orgaddress{\street{Königstuhl 17}, \city{Heidelberg}, \postcode{69117}, \country{Germany}}}

\affil[7]{\orgdiv{Laboratoire d'Astrophysique de Marseille}, \orgname{Aix Marseille Univ, CNRS, CNES},\orgaddress{ \city{Marseille}, \country{France}}}


\abstract{Brown dwarfs are massive analogues of extrasolar giant planets. Compared to exoplanets whose observations are generally limited by the presence of their bright host star, brown dwarfs are ideal targets for studying substellar atmospheric physics, chemistry and dynamics.
Previous observations and simulations of their atmospheres suggest preferential cloud formation around the equator, associated with an equator-pole thermal gradient. Here we show that this atmospheric structure should induce latitudinal chemical variations detectable by the Doppler effect. We introduce a new method - Differential Molecular Rotational Broadening - which consists in comparing the apparent rotational broadening of individual molecules from high-resolution spectra. Application of this approach to VLT-CRIRES observations for different molecules (CO, H$_2$O, CH$_4$ and NH$_3$) in the atmosphere of the brown dwarf DENIS J0255-4700 confirms the existence of latitudinal chemical variations. Our data suggest a depletion of CH$_4$ and NH$_3$ at low latitudes, consistent with an equatorial cloud belt. Our method could be applied to multiple brown dwarfs and exoplanets to map their atmospheres and to study various atmospheric processes.

}




\maketitle

\section{Introduction}\label{sec1}

Brown dwarfs (objects with masses 13 to 72 times that of Jupiter) offer a unique laboratory for studying substellar atmospheric physics, chemistry and dynamics \cite{showman_atmospheric_2020}. As they cool over time, they experience a wide range of temperatures with transitions of their colors and fluxes in the optical and infrared, in particular the L-T transition (effective temperature $\sim$ 1200K), associated with the shift from CO-dominated atmospheres to CH$_4$-dominated atmospheres. The later is likely related to the breakup or the disappearance of silicate and iron clouds at the photospheric level over the transition from L dwarfs to T dwarfs \cite{kirkpatrick_new_2005, cushing_atmospheric_2008, marley_patchy_2010, charnay_self-consistent_2018}.
The atmospheric circulation of brown dwarfs is driven by a combination of rotation (the rotation periods are generally only a few hours \cite{metchev_weather_2015}), radiative cooling, atmospheric waves and convection triggered by internal heat and clouds \cite{showman_atmospheric_2013, zhang_atmospheric_2014, showman_atmospheric_2019, tan_atmospheric_2021-1, tan_atmospheric_2021, teinturier_clouds_2026}.
Observations of brown dwarf atmospheres have revealed the presence of fast eastward winds (superrotation) and photometric variability, likely related to inhomogeneous cloud covers \cite{allers_measurement_2020, apai_zones_2017, biller_time_2017, metchev_weather_2015, crossfield_global_2014}. 
All these processes are also expected to occur on young giant exoplanets. Brown dwarfs can be used as analogues of these exoplanets observed by direct imaging, with a much higher level of precision.

\section{Prediction from 3D models: latitudinal variations in cloudiness, temperature and chemistry}\label{sec2}

A major prediction from 3D simulations of cloudy L-type brown dwarfs is the preferential cloud formation at low latitudes near the equator \cite{tan_atmospheric_2021, tan_large-amplitude_2025, teinturier_clouds_2026}. Indeed, at high latitudes, the strong Coriolis force for fast rotating objects tends to suppress the generation of planetary waves (large-scale patterns of motion in the atmosphere), and to weaken vertical mixing, limiting cloud formation \cite{tan_atmospheric_2021-1,tan_atmospheric_2021}. The radiative effects of clouds, which warm the atmosphere and trigger convection, tend to amplify this cloud latitudinal variation \cite{teinturier_clouds_2026}.
This outcome from 3D models is consistent with observations of brown dwarfs with high inclinations (equator-view), that appear redder and more variable than those at low inclinations (polar view), as if brown dwarfs were cloudier at the equator than at the poles \cite{vos_viewing_2017, vos_spitzer_2020, suarez_ultracool_2023}. 3D models also predict that the equatorial cloud belt should induce a maximum of temperature at low latitude due to the cloud greenhouse effect \cite{tan_atmospheric_2021, teinturier_clouds_2026}. 
This latitudinal thermal gradient should impact the chemical composition. In particular, CO and N$_2$ are more stable at high temperatures while H$_2$O, CH$_4$ and NH$_3$ are more stable at low temperatures \cite{zahnle_methane_2014, fortney_beyond_2020}. 
3D simulations of brown dwarfs including chemistry show that the chemical composition follows the temperature field at the pressure where vertical quenching occurs. Above this level (at $\sim 10$ bars for the CH$_4$-CO and NH$_3$-N$_2$ conversions, see Extended Figure 1), the vertical mixing is faster than chemical kinetics. Chemical abundances are then efficiently vertically mixed, with a limited impact from the horizontal transport \cite{lee_dynamically_2024}. That can be explained by a comparison of the characteristic mixing timescales. Based on 3D simulations \cite{teinturier_clouds_2026}, the estimated latitudinal mixing timescale ($\tau^{mixing}_{lat}\sim \frac{R_p}{V}$, where $R_p$ is the brown dwarf radius and $V\sim100$ m/s is the mean meridional wind) is much larger than the typical vertical mixing timescale ($\tau^{mixing}_{vert}\sim \frac{H^2}{K_{zz}}$, where $H$ is the atmospheric scale height and $K_{zz}$ is the vertical eddy mixing coefficient, typically of the order of 10$^7$-10$^9$ cm$^2$/s for L and T dwarfs \cite{charnay_self-consistent_2018}). This implies that the atmospheric composition of brown dwarfs should be mostly fixed locally by vertical quenching. 
Figure 1 shows maps of cloud and temperature from a 3D simulation of a brown dwarf at the L-T transition, together with maps of mixing ratios of CO, H$_2$O, CH$_4$ and NH$_3$ derived from vertical quenching (see Methods). At 10 bars (close to the quenching level for the CH$_4$-CO and NH$_3$-N$_2$ conversions), there is a difference of $\sim$200 K between the equator and the poles (see also Extended Figure A1). This creates a depletion of CH$_4$ (factor $\sim$5), H$_2$O (factor $\sim$1.2) and NH$_3$ (factor $\sim$1.5), and an enrichment of CO (factor $\sim$1.1) at low latitudes ($<20^\circ$) compared to high latitudes ($>20^\circ$, see also Extended Figure A2). For L and T dwarfs, the emission in the K band is expected to come mostly from pressures around 0.1-1 bar \cite{morley_spectral_2014, charnay_self-consistent_2018}, so above the quenching level of the CH$_4$-CO and NH$_3$-N$_2$ conversions. The latitudinal chemical variations induced by equatorial clouds should therefore affect emission spectra.

\section{Probing latitudinal chemical variations by Differential Molecular Rotational Broadening}\label{sec3}

Here, we introduce the method of Differential Molecular Rotational Broadening (DMRB) to probe latitudinal chemical variations. It consists in measuring and comparing the rotational broadening (i.e. the projected equatorial velocity $v\mathrm{sin}i$) of different molecules from observations at high-spectral resolution. This method should be efficient on brown dwarfs, given that 
they are generally fast rotators with high rotational broadening.
Figure 2 illustrates the distribution of clouds and species with latitude as well as the line rotational broadening for a depleted equatorial region. An excellent fit to the rotational broadening for a brown dwarf with a relatively high viewing inclination and a depleted band between latitudes $\pm \alpha$ corresponds to (see Methods): 
\begin{equation}
v\mathrm{sin}i = v\mathrm{sin}i_0 \times \sqrt{(1-(\mathrm{sin}i \times\mathrm{sin}\alpha)^{3/2})}
\end{equation}
Where $v\mathrm{sin}i_0$ corresponds to the rotational broadening for a homogeneous photosphere with no equatorial depletion ($\alpha=0$) \cite{gray_observation_2005}. For a high inclination with $\mathrm{sin}i \sim 1$, $\alpha$ is directly linked to the measured $v\mathrm{sin}i$.
A depletion between $\pm 15^\circ$ (similar to CH$_4$ in Figure 1) should therefore lead to a reduction of the measured $v\mathrm{sin}i$ by $\sim$7$\%$. Of course, it corresponds to an idealized case since the depletion in the equatorial band should not be total (see Figure 1), limiting the change of apparent $v\mathrm{sin}i$ (see Extended Figure A3).
The uncertainty on the measurement of $v\mathrm{sin}i$ is typically of the order of $\frac{0.5 \times v\mathrm{sin}i}{ SNR}$, where $SNR$ is the signal-to-noise ratio associated to the detection of a molecule. It means that the detection of chemical variations at more than 3$\sigma$ should require signal-to-noise ratios higher than $\sim$20. Such precisions are now accessible to many sub-stellar objects observed with ground-based spectrographs. Since CO and CH$_4$ are expected to show opposite variations with latitude, brown dwarfs that show clear signatures of both molecules, combined with a high inclination and a fast rotation rate, are ideal targets to probe the chemical coupling to a potential equatorial cloud belt.

\section{Application to the late-L dwarf DENIS J0255-4700}\label{sec4}

We tested the DMRB method on VLT-CRIRES observations of the late L dwarf DENIS J0255-4700 (ICRS coordinates:  RA=02h55m03.6927058871s, DEC=-47${^\circ}$00'51.356903955"). It was observed on November 2, 2022 as part of the
ESO SupJup survey (Program ID: 110.23RW, PI: Snellen).
This brown dwarf is an ideal test case since 1) it has a high rotation rate (global $v\mathrm{sin}i\sim$41 km/s \cite{mohanty_rotation_2003, zapatero_osorio_spectroscopic_2006, de_regt_eso_2024}, ) and likely a high inclination, 2) it is close to the L-T transition, where the thermal gradient induced by clouds should be maximal \cite{teinturier_clouds_2026}, and 3) a previous analysis of CRIRES data revealed clear detections of CO, H$_2$O, CH$_4$ and NH$_3$ with high signal-to-noise ratios \cite{de_regt_eso_2024}. Given the measured radius of $\sim$0.78 R$_{\mathrm{Jup}}$ (consistent with evolutionary models \cite{baraffe_evolutionary_2003}), the rotation period should be less than 2.3 hours ($P=\frac{2\pi R}{v}< \frac{2 \pi R}{v\mathrm{sin}i}$, where $P$ is the rotation period, $R$ is the radius and $v$ is the equatorial velocity). 
Previous analyzes of photometric light curves of DENIS J0255-4700 suggest a rotation period of 1.7-2.2 h with weak variations at longer timescales \cite{morales-calderon_sensitive_2006, koen_ic_2005}. 
To better constrain the rotation period an the inclination, we analyzed TESS data for the sector of DENIS J0255-4700, covering almost 55 days of observations (see Methods). We detected a peak at 2.21 h that we interpret as the rotation period. Based on this measurement, we estimate that the inclination is approximately 72$^{\circ}$.

The total observation exposure time is one hour, which means that the integrated observation is averaged over approximately half a rotation period or a hemisphere. This ensures that it is representative of the longitudinally averaged state of the atmosphere, thus limiting the impact of any longitudinal inhomogeneities. To measure the rotational broadening of each individual molecules, we performed cross-correlation between observation residuals and model residuals. These residuals represent the difference between the observed spectrum (or the full model spectrum, including all molecules) and a model spectrum that excludes the molecule under consideration (see Methods).
The model spectra correspond to the best fit from a free atmospheric recovery \cite{de_regt_eso_2024} and were calculated with atmospheric model PetitRADTRANS \cite{molliere_petitradtrans_2019, blain_spectralmodel_2024}. 
Our calculation also includes the effect of correlated noise \citep{de_regt_eso_2024}. We combined spectral orders 5 to 7 (2.228-2.472 $\mu$m) from CRIRES observations. They correspond to orders for which at least one molecule in addition to H$_2$O is detected (see Methods and Extended Figure A6). This selection is done to compare the rotational broadening in the same conditions, limiting biases.

Figure 3 shows the observed spectrum and residuals for CH$_4$ in the sixth spectral order (see Extended Figure 4 for the other molecules). The excellent fit between observation and model residuals shows how efficiently the contribution of individual molecules can be extracted. 
Figure 4 shows maps of the signal-to-noise ratio (SNR), and maps of the variations of log-likelihood ($\Delta \log L$) \cite{zucker_cross-correlation_2003} compared to to the maximum of log-likelihood, for each molecule. These maps of SNR and $\Delta \log L$ are based on projected rotational velocities ($v\mathrm{sin}i$) and radial velocity (RV). With our formulation of the log-likelihood, the peaks of SNR and $\log L$ are equivalent. 
The cross-correlation and auto-correlation functions (CCF, ACF) are also compared at the peak of SNR in Figure 4. The excellent agreement between the CCF and ACF spectra for all cases except NH$_3$ demonstrates once again how well these model spectra reproduce the observations, allowing the rotational broadening associated with each molecule to be correctly extracted.

Figure 5 summarizes the measured $v\mathrm{sin}i$. We used the variations of the log-likelihood to estimate uncertainties for each measurement (see Methods). 
We measured $v\mathrm{sin}i(\mathrm{all})=42.1_{-0.13}^{+0.13}$ km/s for all molecules (relatively similar to a previous estimation of $41.05_{-0.19}^{+0.19}$ km/s \cite{de_regt_eso_2024}), $v\mathrm{sin}i(\mathrm{CO})=41.2_{-0.19}^{+0.19}$ km/s, $v\mathrm{sin}i(\mathrm{H_2O})=41.5_{-0.18}^{+0.18}$ km/s, $v\mathrm{sin}i(\mathrm{CH_4})=37.0_{-0.40}^{+0.40}$ km/s and $v\mathrm{sin}i(\mathrm{NH_3})=31.8_{-1.48}^{+1.48}$ km/s. 

We therefore obtain $ ( v\mathrm{sin}i(\mathrm{CO}) , v\mathrm{sin}i(\mathrm{H_2O})) > (v\mathrm{sin}i(\mathrm{CH_4}) , v\mathrm{sin}i(\mathrm{NH_3}))$. The variations of $ v\mathrm{sin}i$ are consistent with predictions from 3D simulations of a warm equatorial belt. CH$_4$, whose abundance is expected to show the greatest latitudinal contrast, is probably the best proxy for the latitudinal extension of the cloud belt.
The $ v\mathrm{sin}i$ value of CH$_4$ corresponds to 
an equatorial depletion between latitudes $\pm20^\circ$, relatively similar to our 3D simulations ($\pm15^\circ$).  We conclude that DENIS J0255-4700 possesses an equatorial cloud belt between approximately latitudes $\pm20^\circ$. 

We note that the predicted equatorial depletion of NH$_3$ is relatively smaller than that of CH$_4$. A lower $v\mathrm{sin}i$ for NH$_3$ could be obtained due to its low abundance, producing weak absorbing lines in emission spectra. The best-fitting $v\mathrm{sin}i$ for NH$_3$ could also be biased by the small offset in radial velocity (see Figure 4 and \cite{de_regt_eso_2024}) potentially linked to uncertainties in the wavelengths of NH$_3$ lines from databases. By analyzing the data order by order and exposure by exposure, or by changing the low frequency filtering,
the measured values of $v\mathrm{sin}i$ are only slightly changed (see Extended Table 1 and Figure A6). 
We conclude that these chemical variations with latitude are robust. The contrast of $v\mathrm{sin}i$ between CH$_4$ and H$_2$O is higher than expected from the 3D simulation. However, this simulation was performed for a rotation period of 5 hours. A smaller rotation period should lead to stronger cloud/thermal latitudinal variations \cite{tan_atmospheric_2021}. Other effects could enhance the contrast of $v\mathrm{sin}i$. A prograde equatorial jet would increase rotational broadening at low latitudes (with a weaker effect for CH$_4$ and NH$_3$), with a change of the order of 0.5 km/s based on observations and simulations \cite{allers_measurement_2020, teinturier_clouds_2026}. Similarly, oblateness would increase the variations of $v\mathrm{sin}i$. The impact should be limited for field brown dwarfs (difference between equatorial radius and polar radius lower than 5$\%$), but it could be significant for low-gravity brown dwarfs and young giant exoplanets (difference between equatorial radius and polar radius of 5-30$\%$) \cite{sanghavi_photopolarimetric_2018}.
A consequence of our finding is that some previous measurements of $v\mathrm{sin}i$ and consequently of inclination could be biased by latitudinal chemical variations. Accurate measurements of overall rotation ideally require targeting species with homogeneous distributions.

\section{Perspectives with current and future instruments for brown dwarfs and exoplanets}\label{sec5}

Doppler imaging is a technique to map stellar surface, often based on the hypothesis of the maximum entropy for image reconstruction. It is the ideal technique to map the atmospheres of brown dwarfs and to reveal horizontal variations in cloudiness, temperature, and chemical composition \cite{vogt_doppler_1987, crossfield_global_2014}. However, this method requires a high signal-to-noise ratio and time-resolved observations, strongly limiting its applicability. Moreover, this method requires a complicated algorithm, and it is less sensitive to latitudinal variations than to longitudinal variations \cite{chen_global_2024}.
Until now, Doppler imaging has been applied to only the nearest brown dwarfs: Luhman 16 A and Luhman 16 B (WISE 1049-5319 A and B), without isolating the different molecular contributions \cite{crossfield_global_2014, chen_global_2024}. DMRB is a complementary method to Doppler imaging, since it is more sensitive to latitudinal variations. It is also significantly simpler, as it does not rely on time-resolved observations and does not require a full image reconstruction to reveal latitudinal variations. Longitudinal variations are averaged by observing a significant fraction of a rotational period or by stacking observations at different epochs. By scaling our results with a catalog of L and T dwarfs, we estimate that 20 brown dwarfs are accessible for detecting latitudinal variations for 1-hour observations with a 8 m telescope (see Methods and Extended Figure A8). This number rises up to 100 objects for 5-hour observations, and up to 300 objects for 1-hour observations with a 39 m telescope (ELT).
DMRB can thus be applied to a large number of objects with current instruments (e.g. VLT-CRIRES, CFHT-SPIRou). Our analysis reveals the richness of high-resolution spectra but it remains qualitative. To be more quantitative, a full atmospheric retrieval analysis (i.e. inverting from the observed spectra the vertical profiles of temperature, chemical abundances and clouds for multiple latitudinal bands) would be necessary to reconstruct accurate latitudinal maps. Additional 3D simulations, fully coupling clouds and chemistry, would also be required to investigate more precisely the chemical quenching in the atmospheres of brown dwarfs and young giant exoplanets.

The technique presented here has great potential for studying various atmospheric processes. In addition to probing chemical variations, it could be used to map clouds. Indeed, if the cloud layer varies with latitude and masks thermal emission in a spectral window, the measurement of $v\mathrm{sin}i$ for a given molecule (e.g. H$_2$O) at different spectral bands could constrain the altitude of the cloud layer as a function of latitude. In particular, for an equatorial cloud belt, $v\mathrm{sin}i$ should be smaller in the J band (probing deep in the atmosphere, where clouds are optically thick) than in the K band (probing above the cloud top).
This technique could also be used to investigate auroral processes, such as the auroral emission by H$_3$+, or the methane emission detected on an isolated Y dwarf \cite{faherty_methane_2024}. These processes should occur at high latitudes and the corresponding emission lines should thus be associated with a low rotational broadening. Similarly, photochemistry and auroral chemistry can induce strong latitudinal chemical variations, as it is the case in Jupiter's stratosphere \cite{rodriguez-ovalle_stratospheric_2024, rodriguez-ovalle_jwst_2025}.
Finally, using the DMRB technique with the next generation of spectrographs (e.g. instruments ANDES, METIS and PCS on the ELT \cite{palle_ground-breaking_2025, brandl_instrument_2010, kasper_pcs_2021}) could allow us to map the atmospheres of exoplanets observed in thermal emission, in reflected light, but also in transmission, for fast rotating objects.
For instance, using a simple contrast scaling with Proxima b \cite{palle_ground-breaking_2025}, ELT-ANDES could detect H$_2$O in reflected light
on the temperate giant planets GJ 876 b $\&$ c, with a SNR higher than 30 in one night. This detection level, should be sufficient to search for latitudinal variations. Detailed simulations will be essential to fully assess the potential of DMRB for upcoming instruments.

\clearpage

\section{Figures}\label{sec6}

\begin{figure}[h] 
	\centering
	\includegraphics[width=0.45\textwidth]{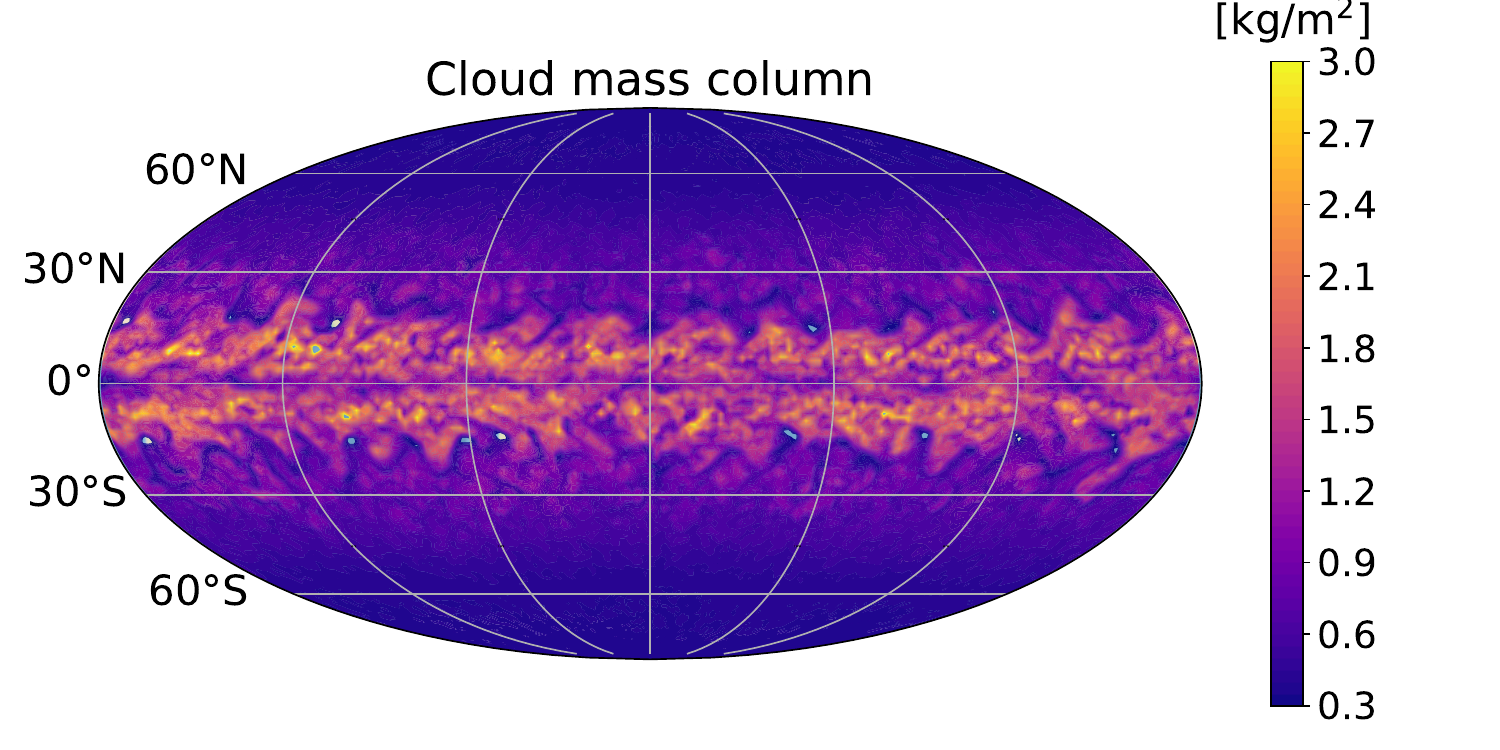} 
    \includegraphics[width=0.45\textwidth]{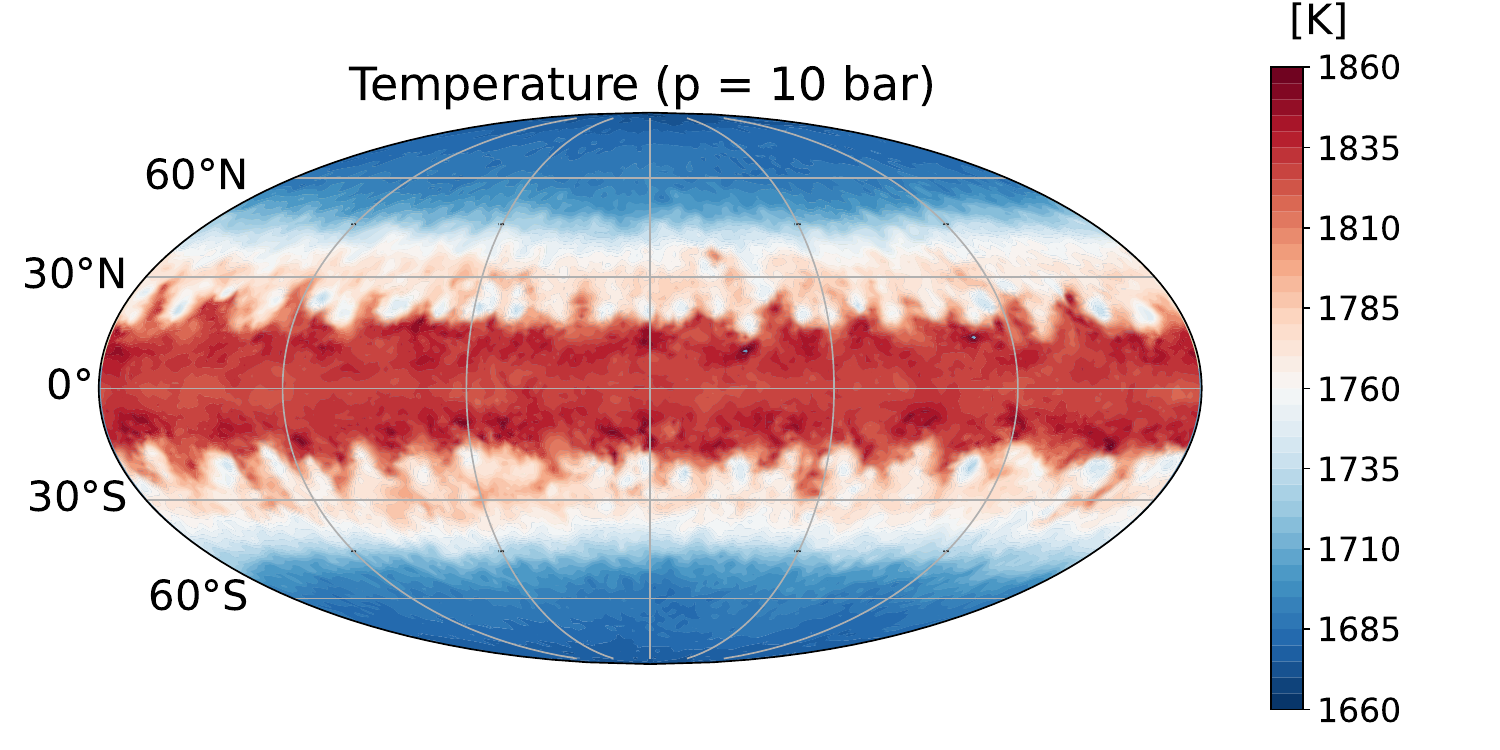}
	\includegraphics[width=0.45\textwidth]{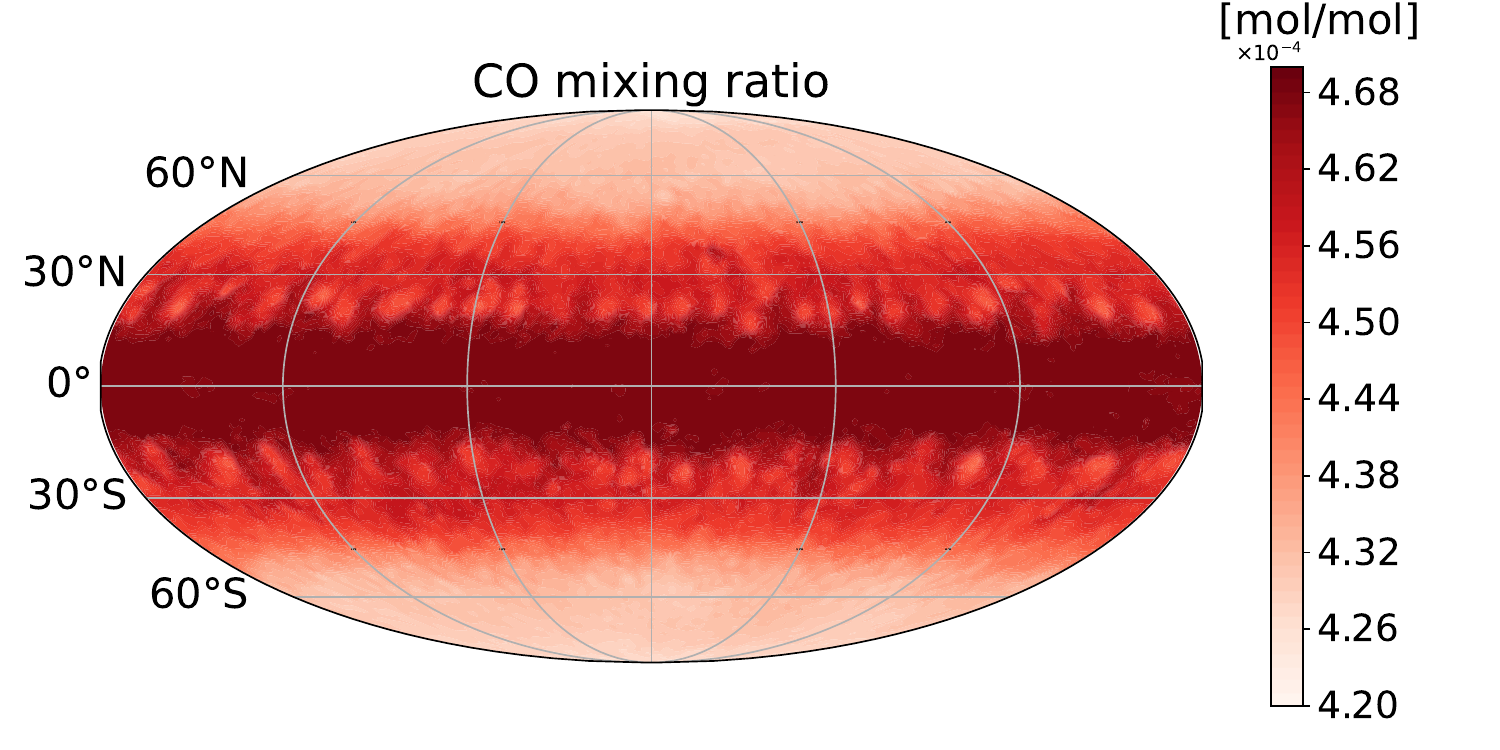}
	\includegraphics[width=0.45\textwidth]{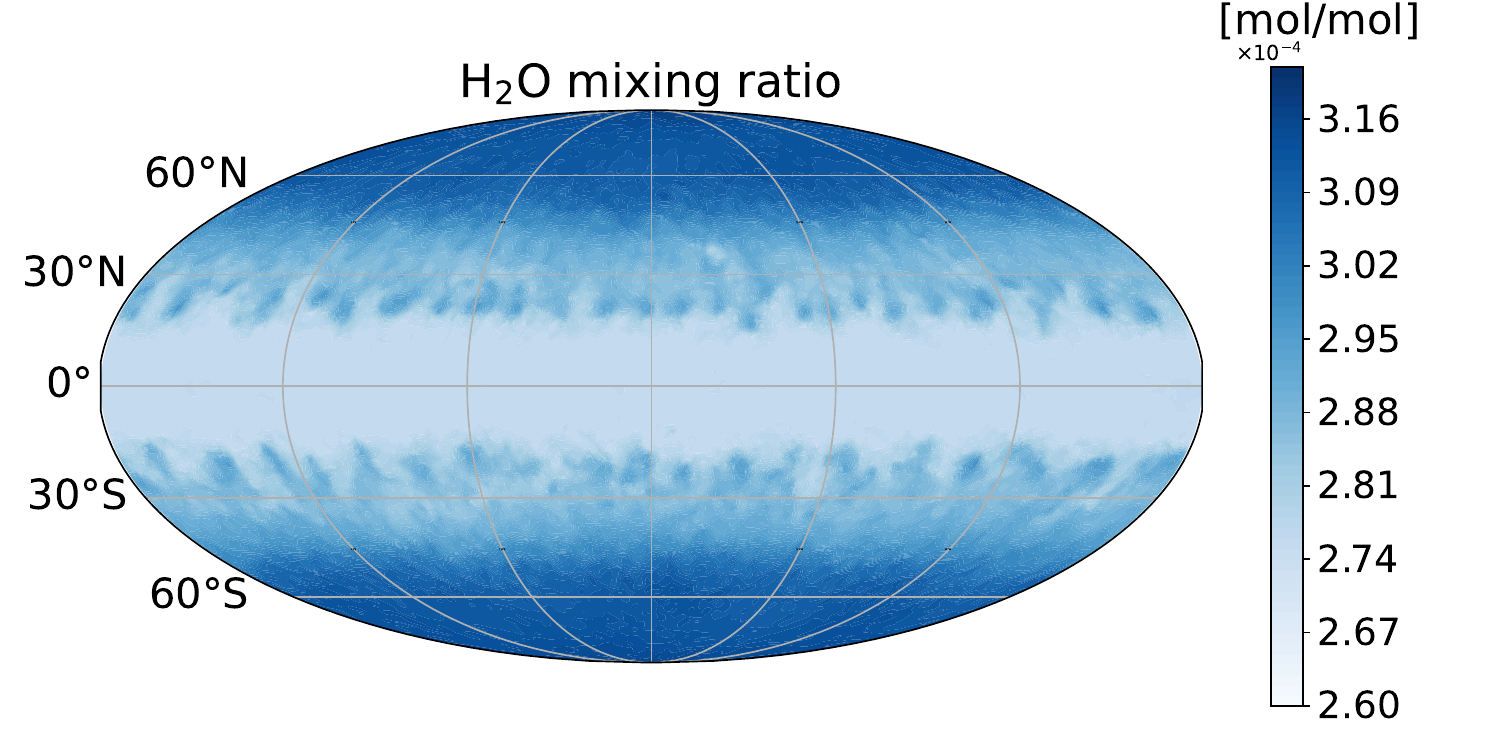} 
	\includegraphics[width=0.45\textwidth]{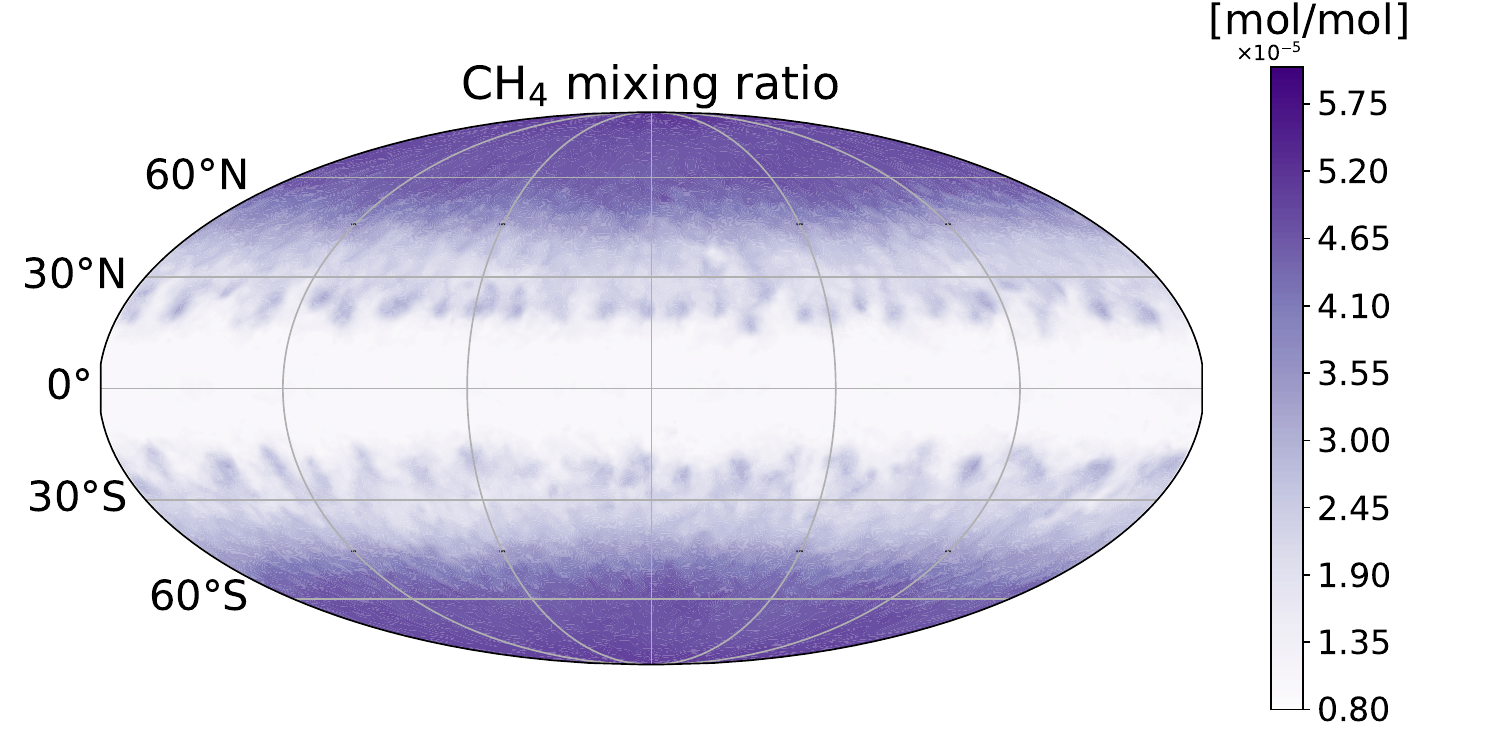} 
	\includegraphics[width=0.45\textwidth]{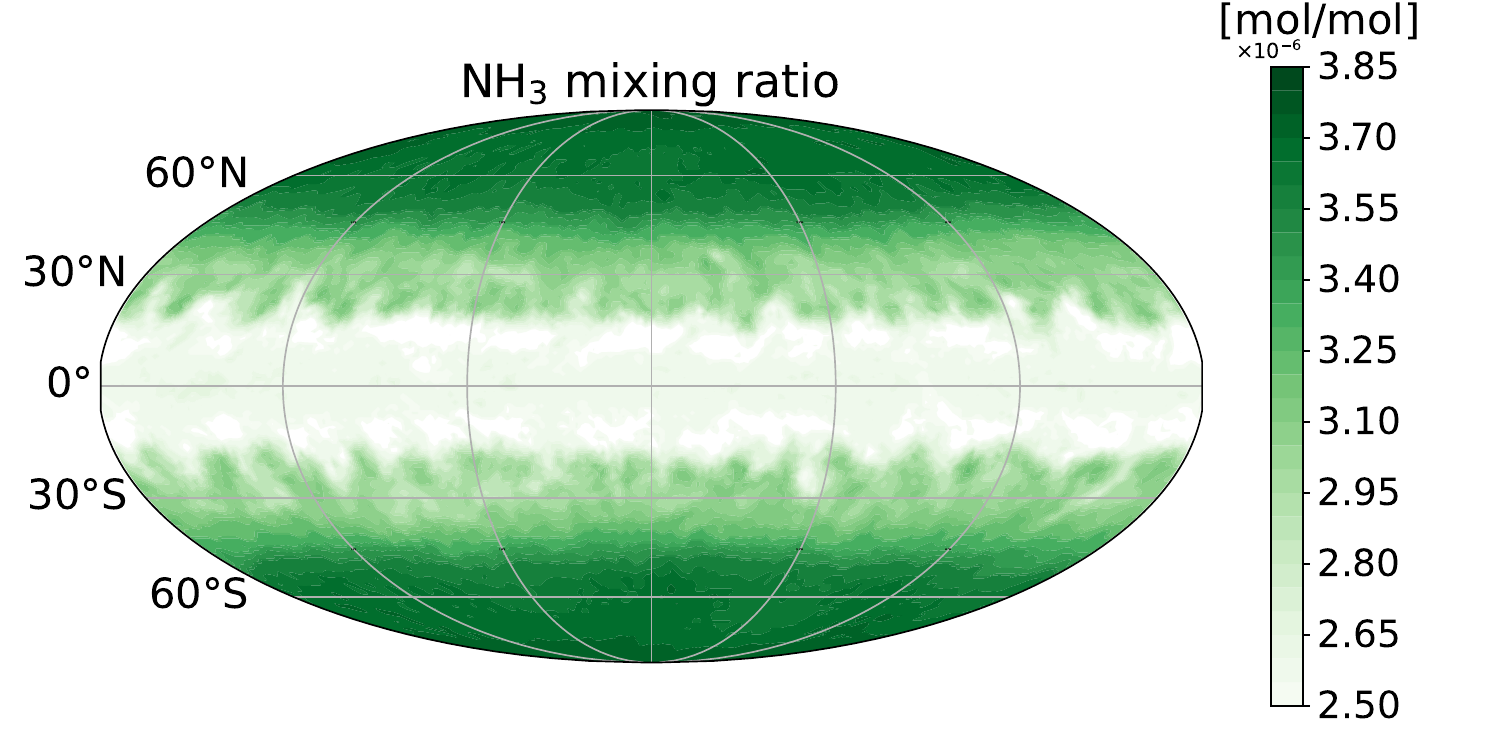}      
	\caption{Maps of cloud mass column (top left), temperature at 10 bars (top right), CO mixing ratio (middle left), H$_2$O mixing ratio (middle right), CH$_4$ mixing ratio (bottom left) and NH$_3$ mixing ratio (bottom right) for a brown dwarf at the L-T transition. The cloud and temperature maps are outputs from a 3D simulation of a brown dwarf (T$_{\mathrm{eff}}$ = 1000 K, rotation period = 5 hours, gravity = 10$^5$ cm/s$^2$, solar metallicity and solar C/O) including silicate clouds (particle radii of 20 $\mu$m) \cite{teinturier_clouds_2026}. The mixing ratio maps were computed from the 3D thermal structure with chemical vertical quenching, using parametrizations of chemical timescales \cite{zahnle_methane_2014} and an eddy diffusion coefficient K$_{\mathrm{zz}}$ = 10$^8$ cm$^2$/s.}
	\label{fig1} 
\end{figure}

\begin{figure}[h] 
	\centering
	\includegraphics[width=0.35\textwidth]{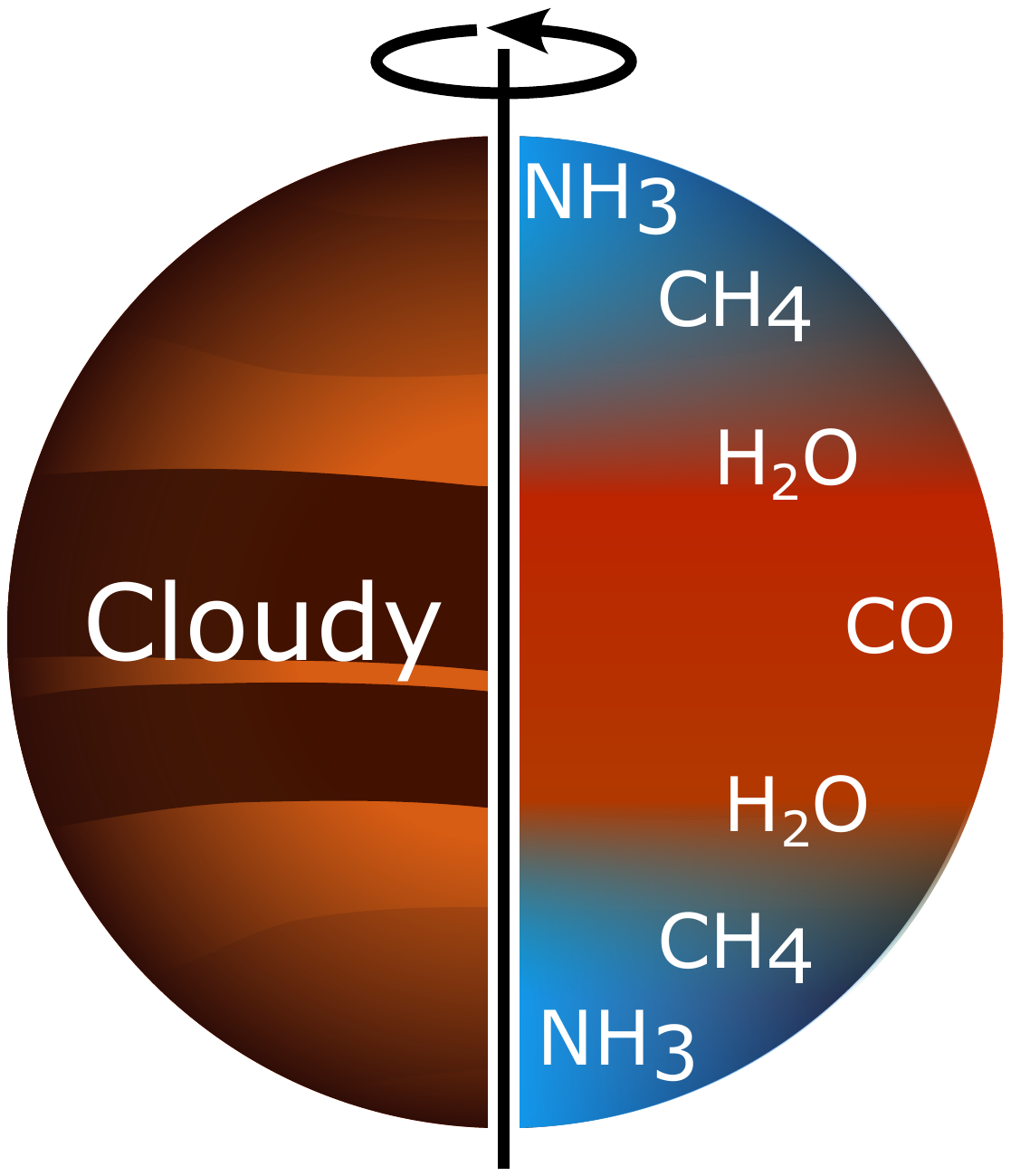} 
	\includegraphics[width=0.55\textwidth]{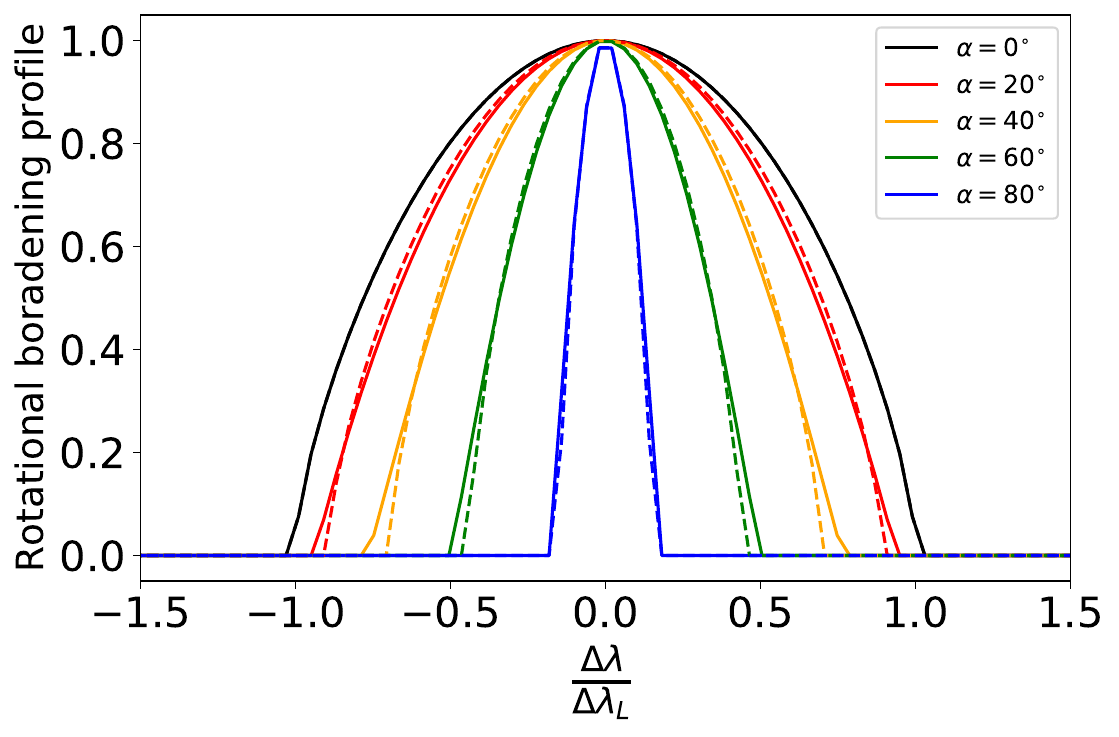} 
	\caption{(A) Illustration of the latitudinal variations of clouds, temperature (warm equatorial regions in red and cold polar regions in blue) and chemical composition. (B) Function of rotational broadening (solid line) assuming a depleted equatorial band between latitudes $\pm \alpha$. $\Delta \lambda$ is the offset in wavelength compared to the line center at $\lambda_L$ and $\Delta \lambda_L = \lambda_L v\mathrm{sin}i/c$ is the maximal Doppler shift for an equatorial velocity $v$ and an inclination $i$. The dashed line is a fit corresponding to the convolution function of rotational broadening for a homogeneous disk \cite{gray_observation_2005} with $\Delta \lambda_L$ multiplied by $\sqrt{(1-\mathrm{sin}(\alpha)^{3/2})}$ (see Methods).
    }
	\label{fig2} 
\end{figure}

\begin{figure}[h] 
	\centering
    \includegraphics[width=1\textwidth]{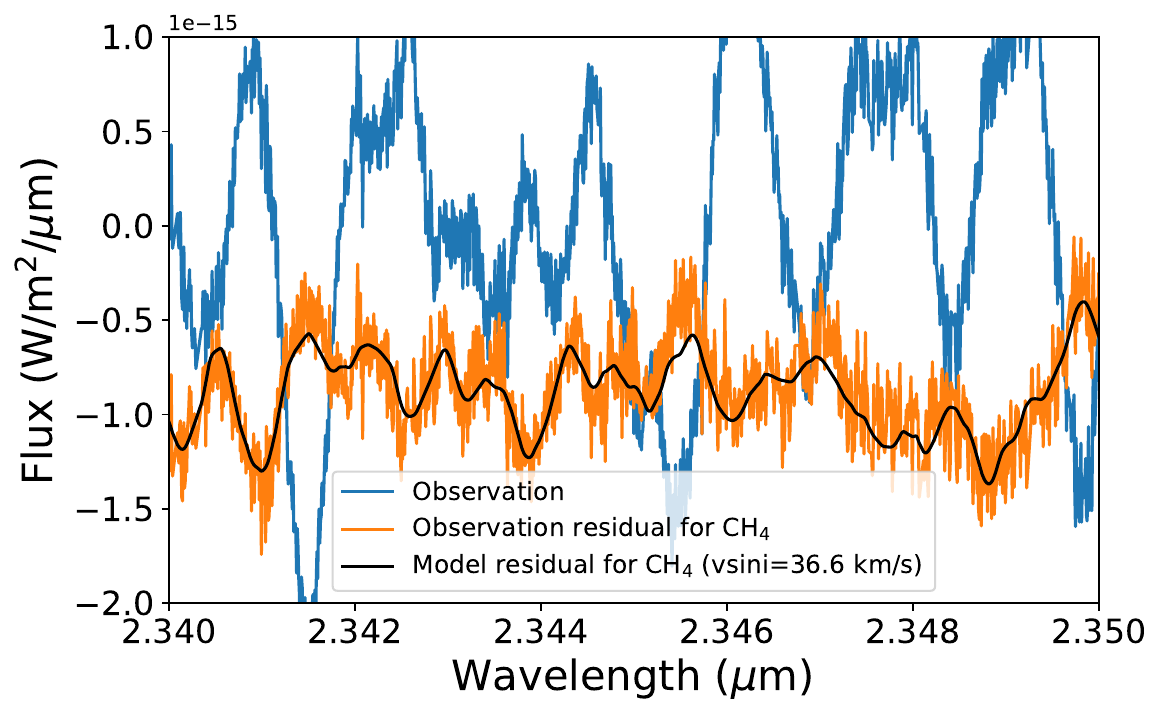} 
	\caption{Spectrum of DENIS J0255-4700 in the sixth spectral order of the CRIRES dataset after data reduction (blue line). The comparison between the observation residual (orange line) and the model residual (green line, with the rotational broadening corresponding to the best fit: $v\mathrm{sin}i=38.2$ km/s) for CH$_4$ is also indicated.}
	\label{fig3} 
\end{figure}

\begin{figure}[h] 
	\centering
	\includegraphics[width=0.32\textwidth]{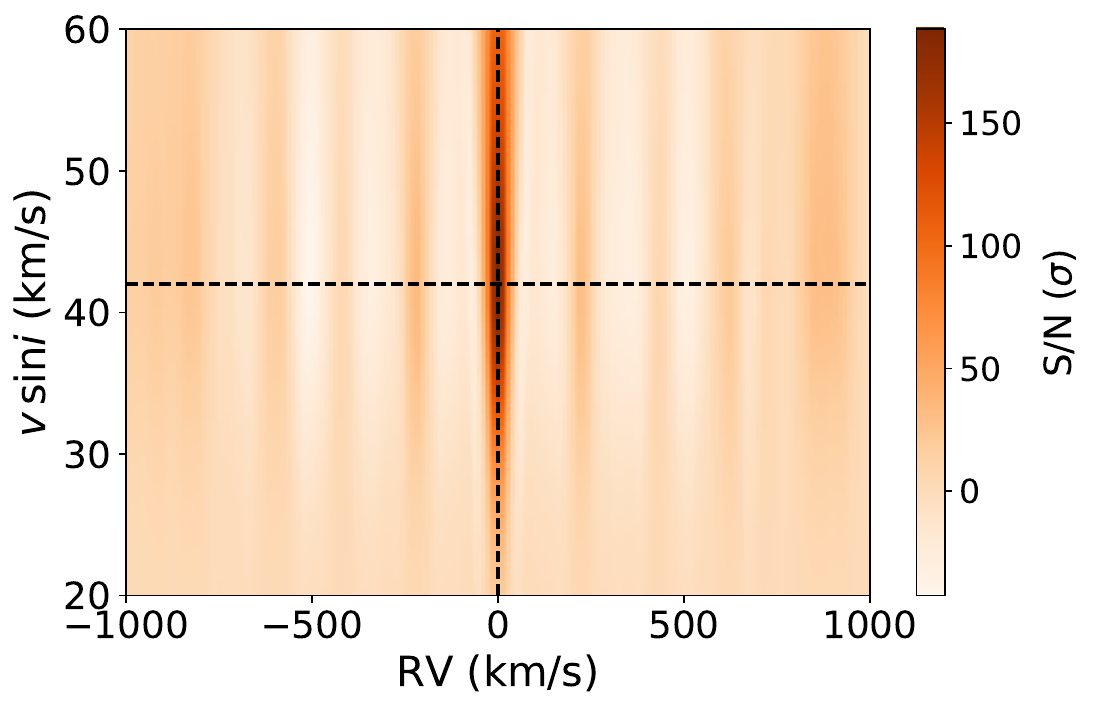} 
    \includegraphics[width=0.335\textwidth]{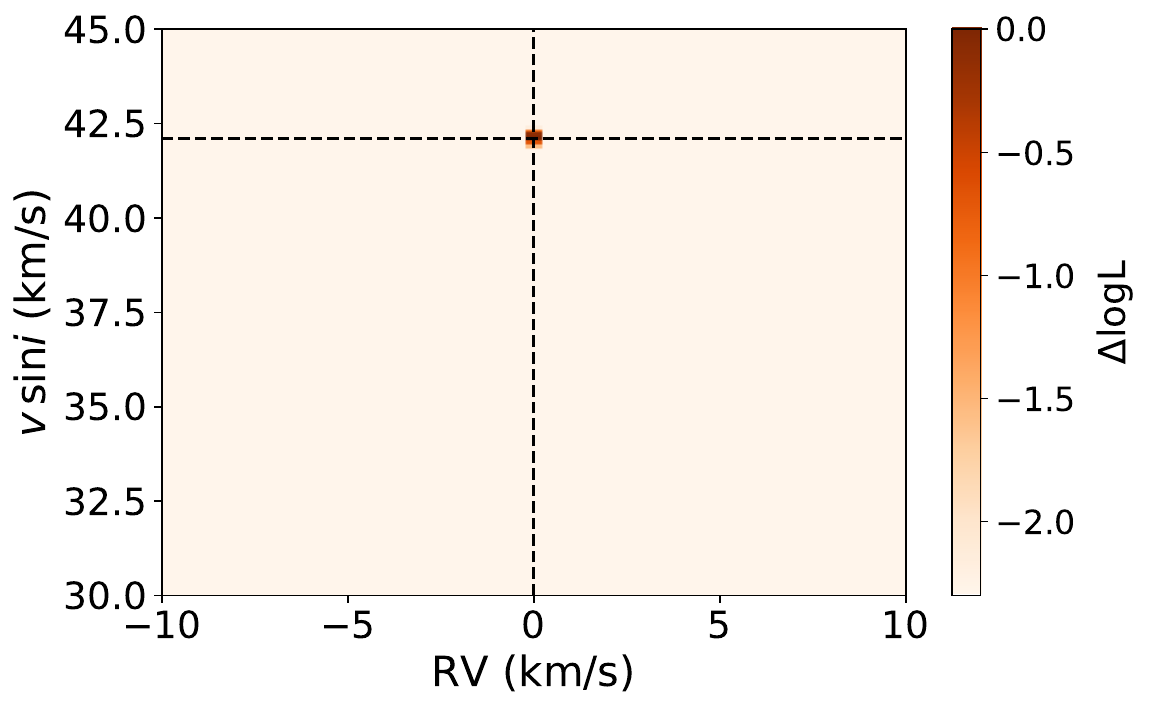} 
    \includegraphics[width=0.33\textwidth]{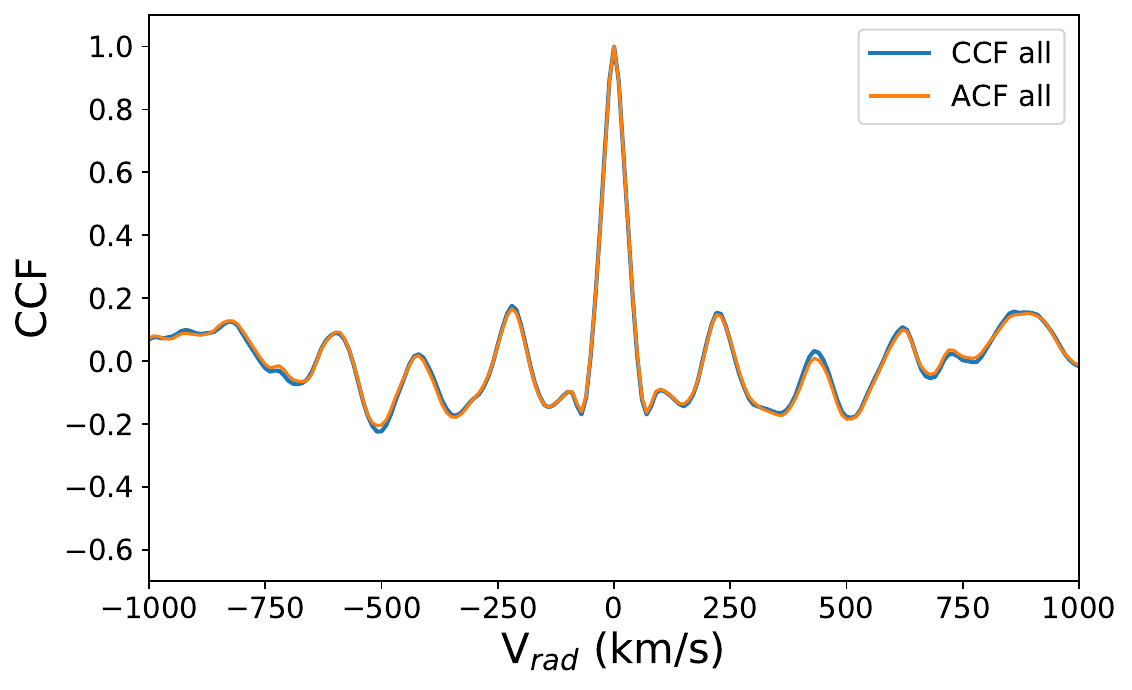}     
	\includegraphics[width=0.32\textwidth]{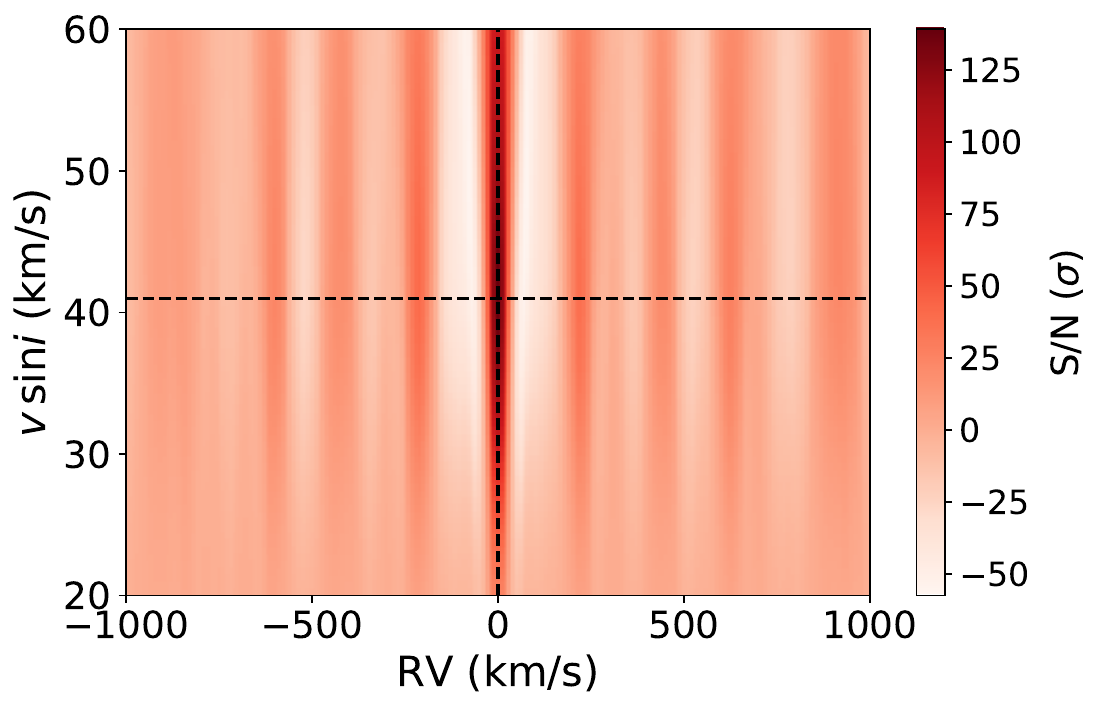} 
    \includegraphics[width=0.335\textwidth]{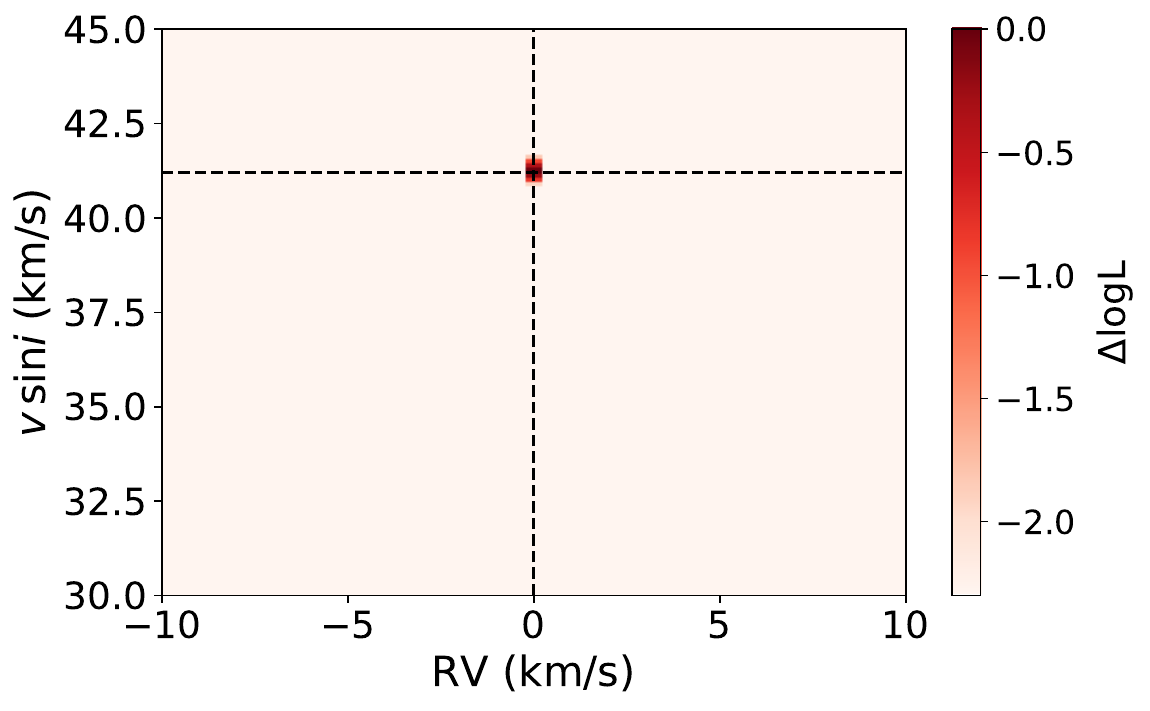} 
    \includegraphics[width=0.33\textwidth]{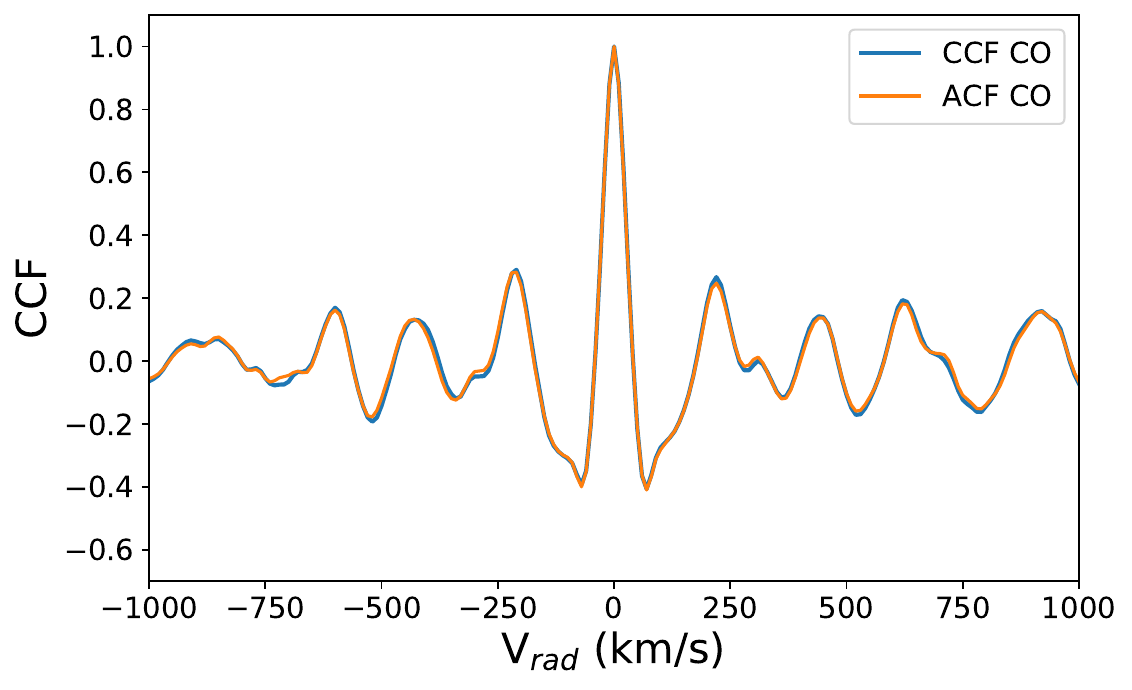}    
	\includegraphics[width=0.32\textwidth]{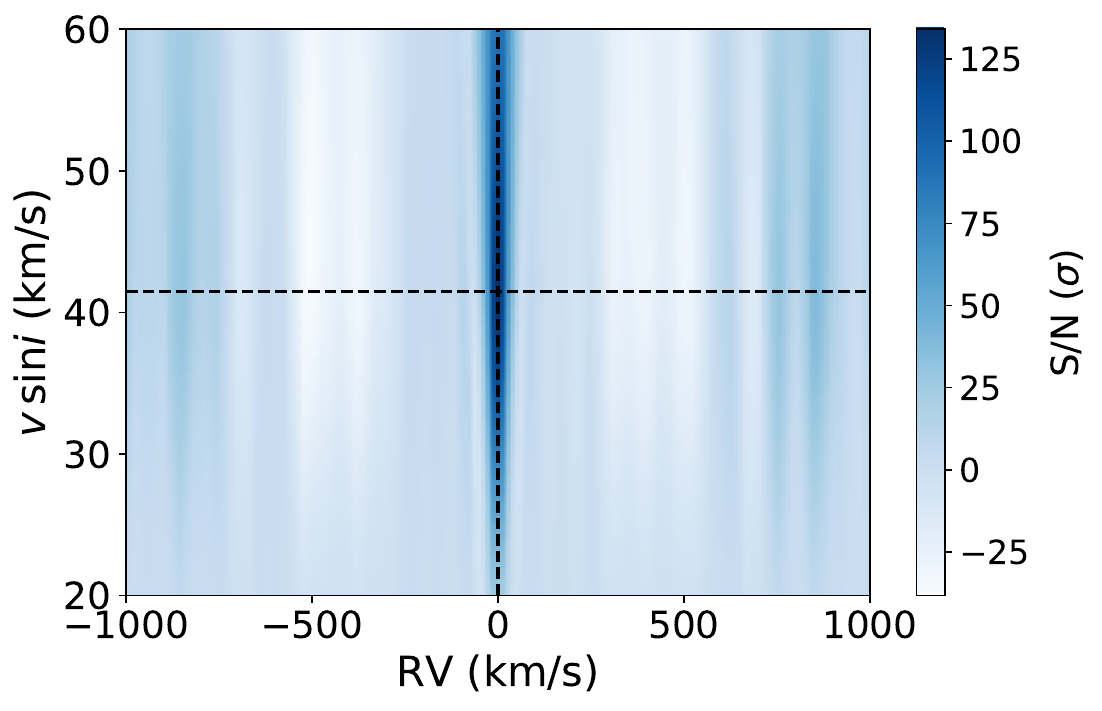}
    \includegraphics[width=0.335\textwidth]{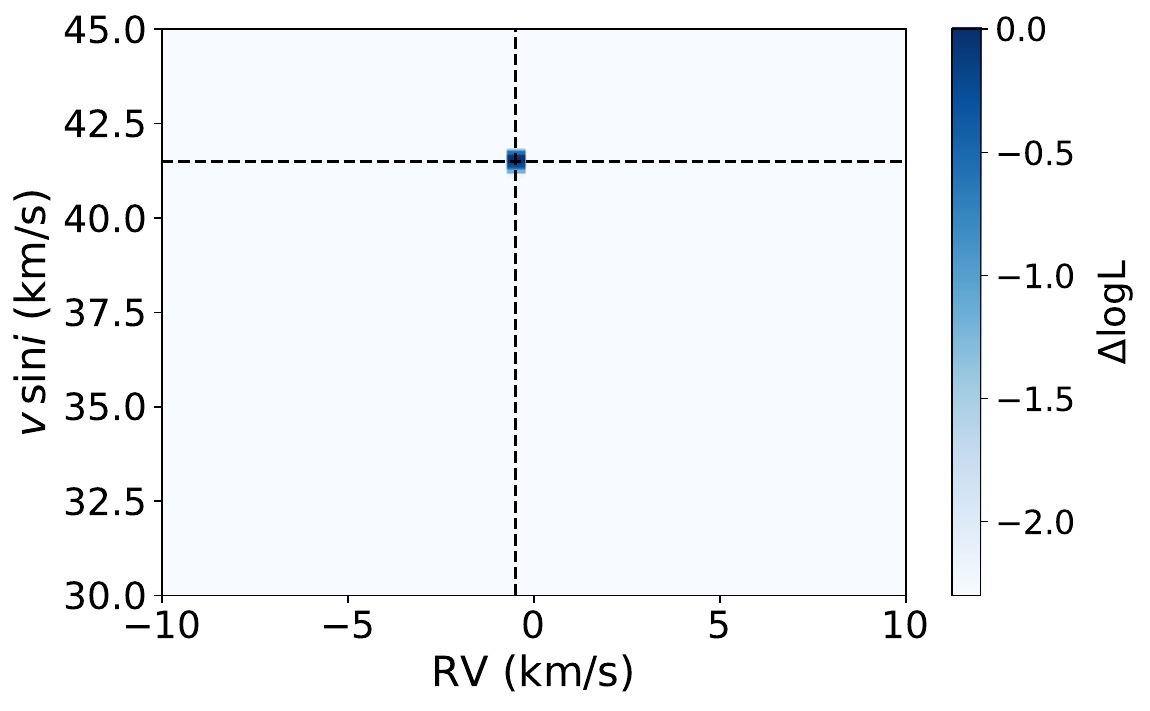} 
    \includegraphics[width=0.33\textwidth]{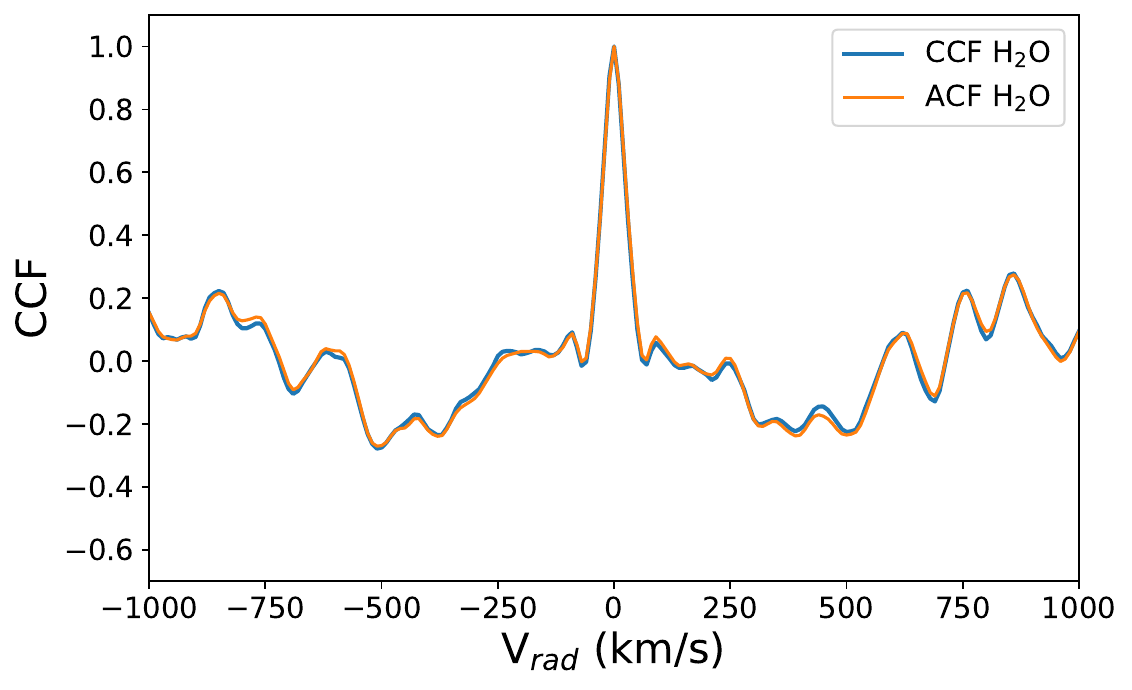}     
	\includegraphics[width=0.32\textwidth]{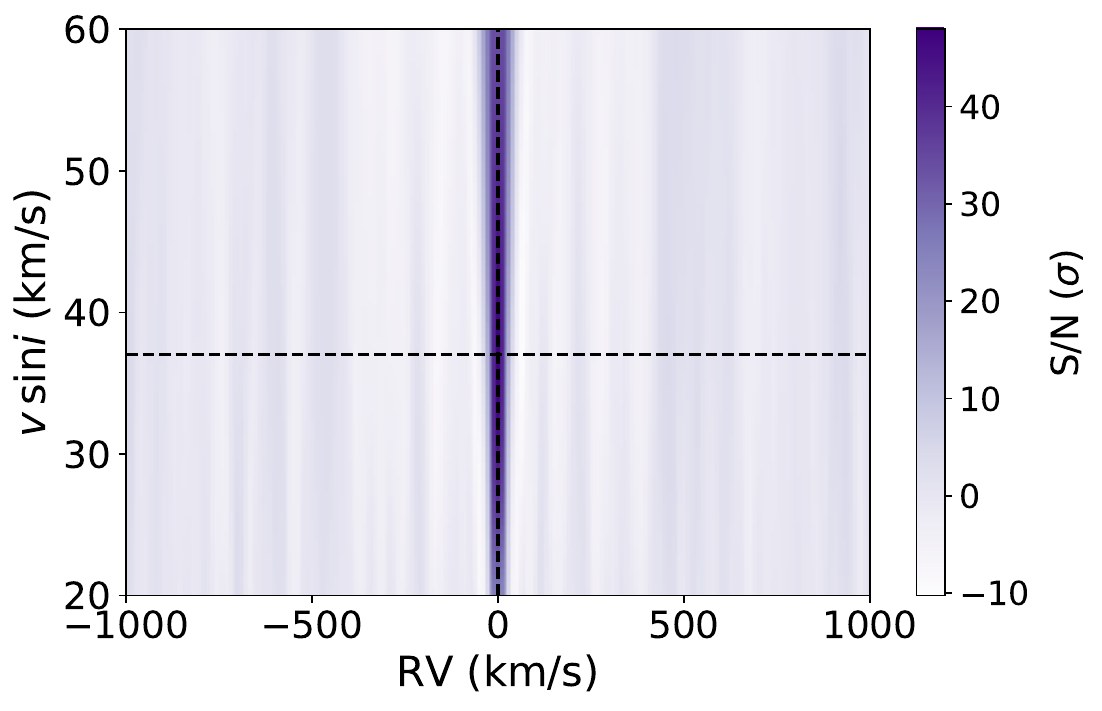} 
    \includegraphics[width=0.335\textwidth]{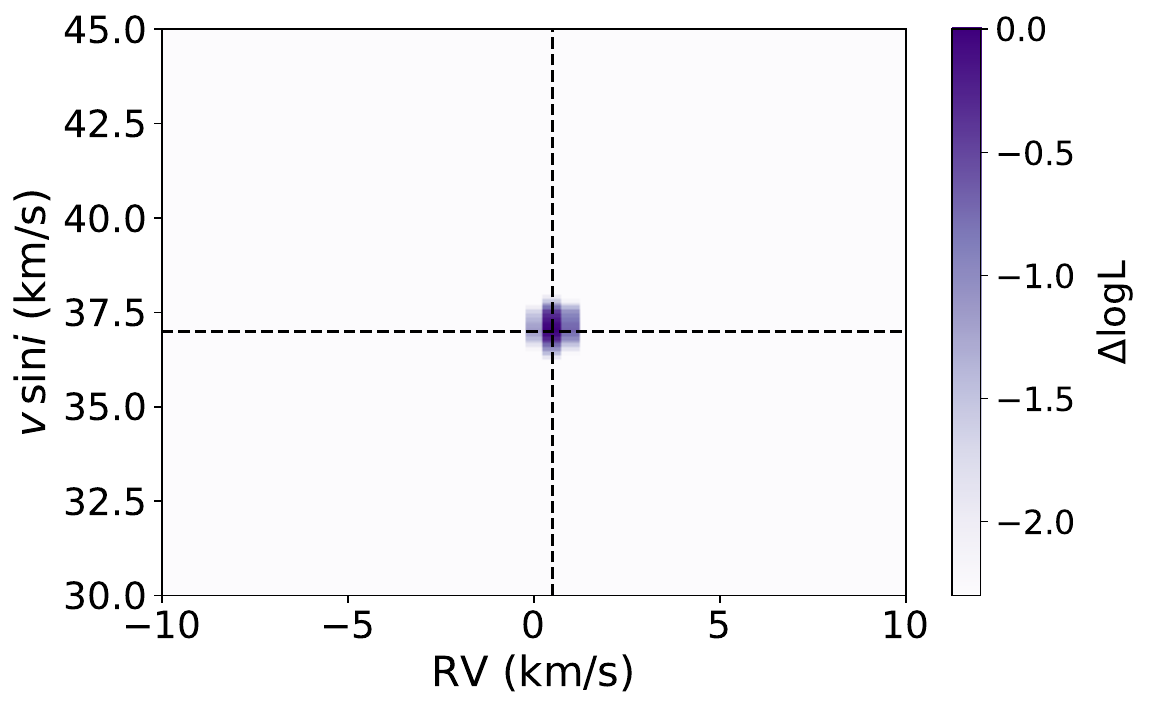} 
    \includegraphics[width=0.33\textwidth]{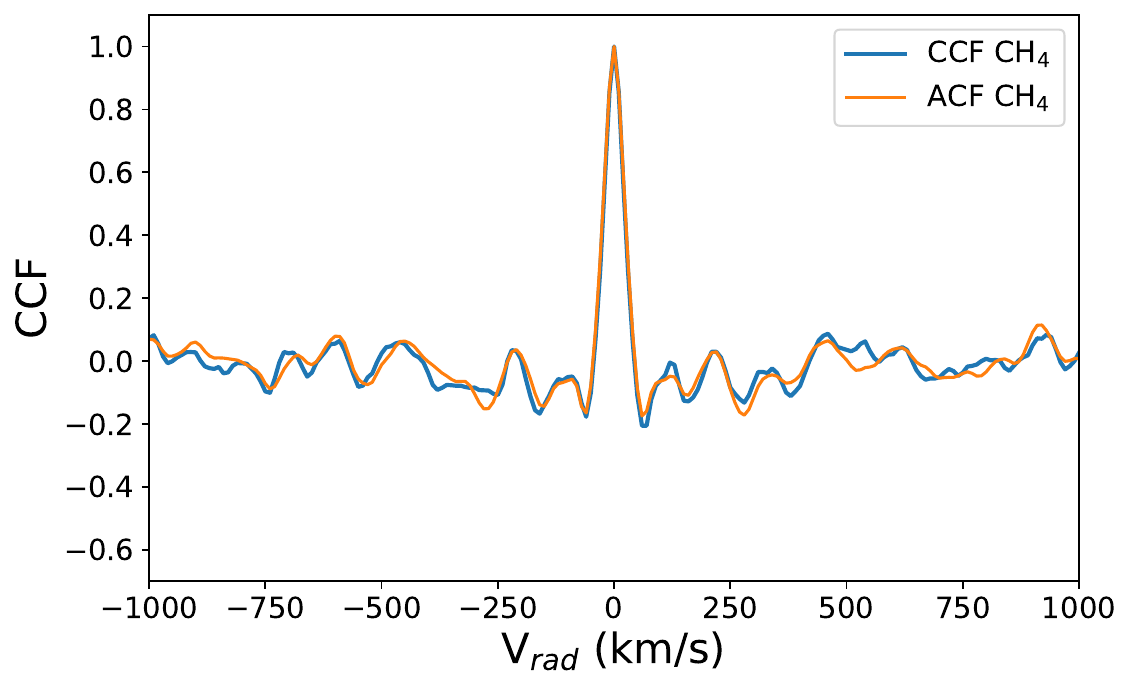}     
	\includegraphics[width=0.32\textwidth]{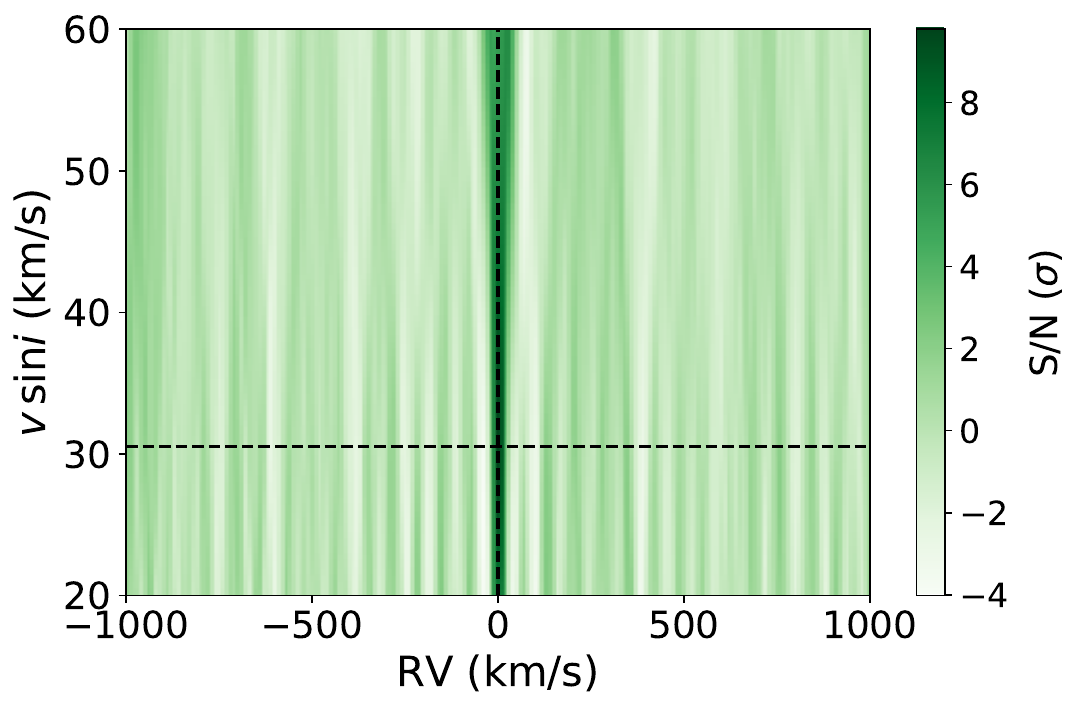} 
    \includegraphics[width=0.335\textwidth]{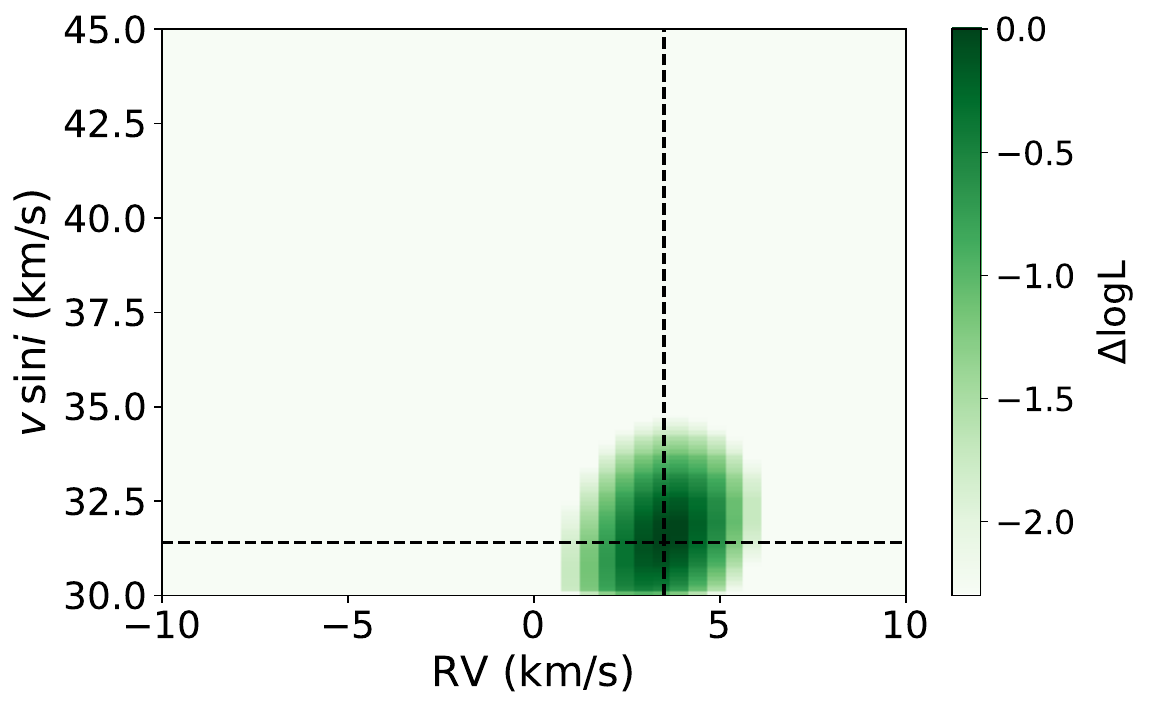} 
    \includegraphics[width=0.33\textwidth]{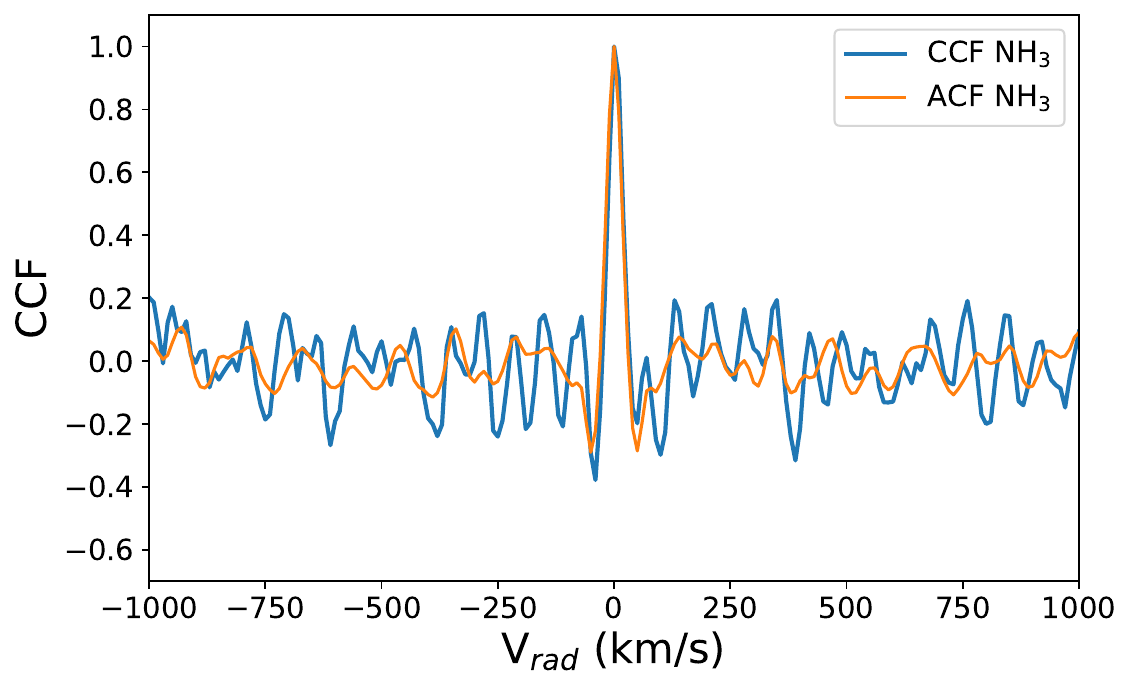}     
 
	\caption{
		(Left) Maps of signal-to-noise ratio (SNR) from the cross-correlation between CRIRES observation of DENIS J0255-4700 and model residuals. The peaks of SNR are indicated with green dashed lines.
        (Middle) Maps of variation of log-likelihood ($\Delta \log L$) compared to the peak of log-likelihood (equivalent to the peak of SNR).
       (Right) Cross-correlations (CCF) and auto-correlations (ACF) at the $v\mathrm{sin}i$ corresponding to the peak of SNR.
        Panels correspond to full spectra (all molecules included), CO, H$_2$O, CH$_4$ and NH$_3$ (from top to bottom).
        }
	\label{fig4} 
\end{figure}

\begin{figure}[h] 
	\centering
	\includegraphics[width=1\textwidth]{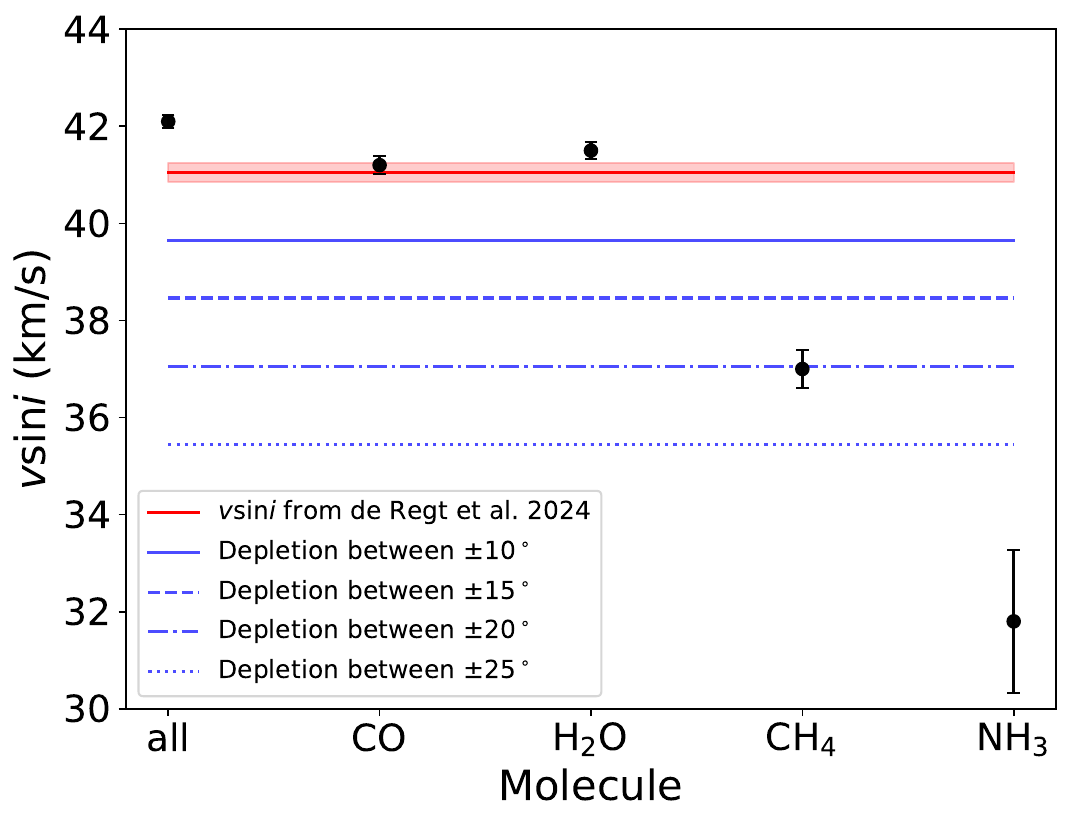} 
	\caption{
		Measured rotational broadening ($v\mathrm{sin}i$) of DENIS J0255-4700 for all molecules, CO, H$_2$O, CH$_4$ and NH$_3$. The red line with the shaded area corresponds to the previous measurement of $v\mathrm{sin}i = 41.05_{-0.19}^{+0.19}$ km/s for the full K-band CRIRES spectrum \cite{de_regt_eso_2024}. The blue lines correspond to values of $v\mathrm{sin}i$ for different equatorial depletions (between latitudes $\pm10^\circ$,  $\pm15^\circ$,  $\pm20^\circ$ and $\pm25^\circ$), using formula (1) with $v\mathrm{sin}i_0 = 41.05$ km/s and $i = 72^{\circ}$.
        }
	\label{fig5} 
\end{figure}

\clearpage

\section{Methods}\label{sec11}

\subsection*{3D simulation of brown dwarf}
Figure \ref{fig1} was calculated from the results of a 3D simulation corresponding to a brown dwarf at the L-T transition \cite{teinturier_clouds_2026}. The simulation was performed with the generic Planetary Climate Model (generic PCM) \cite{wordsworth_gliese_2011, spiga_global_2020, teinturier_radiative_2024}. We used a horizontal resolution of 1.40625$^\circ$ $\times$ 0.9375$^\circ$ in longitude-latitude (equivalent to 256 $\times$ 192 for a regular longitude-latitude grid) and 80 vertical levels, covering pressures from 80 bars to 1 mbar. The simulation used in this study was run with T$_{\mathrm{eff}}$ = 1000 K, solar metallicity, solar C/O, gravity = 10$^5$ cm$^2$/s, rotation period = 5 hours, and including silicate clouds with particle radii of 20 $\mu$m. The effective temperature was chosen to be lower than expected for DENIS J0255-4700 ($\sim$1350 K, \cite{charnay_self-consistent_2018}) because our 3D simulations do not include iron clouds, which cause a delayed L-T transition. In addition, we used a simulation with a surface gravity lower than DENIS J0255-4700 by 0.5 dex. Such a difference is approximately equivalent to an effective temperature decrease of 200 K \citep{charnay_self-consistent_2018}. Our simulation should thus be relatively consistent with DENIS J0255-4700.

\subsection*{Calculation of vertical chemical quenching}
We computed the horizontal variations of chemical composition from the temperature field of the 3D simulation assuming non-equilibrium chemistry. To do that, we determined the quenching level assuming an eddy diffusion coefficient K$_{\mathrm{zz}}$ = 10$^8$ cm$^2$/s and comparing the vertical mixing timescale ($t_{mix}=\mathrm{H}^2/\mathrm{K}_{\mathrm{zz}}$, where $\mathrm{H}$ is the atmospheric scale height) to parameterizations of chemical timescales \cite{zahnle_methane_2014}. For the CO-CH$_4$ system, the net reaction is: 
\begin{equation}
\mathrm{H_2O + CH_4 = CO + 3H_2 }
\end{equation}
and the chemical timescale is:
\begin{equation}
t_{\mathrm{CO}}=\left(\frac{1}{t_{q1}} + \frac{1}{t_{q2}}\right)^{-1}
\end{equation}
where
\begin{equation}
t_{q1}=1.5\times10^{-6}p^{-1}m^{-0.7}\exp(42000/T)
\end{equation}
\begin{equation}
t_{q2}=40 p^{-2}\exp(25000/T)
\end{equation}
In these relations, $T$, $p$ and $m$ are the temperature (in K), the pressure (in bars) and the metallicity. 
The chemical composition at the quenching level is fixed by the chemical equilibrium with the equilibrium constant\cite{zahnle_methane_2014}: 
\begin{equation}
K_{\mathrm{CH_4\cdot CO}} = \frac{p\mathrm{CH_4} \cdot p\mathrm{H_2O}}{p\mathrm{CO}\cdot(p\mathrm{H_2})^3}=5.24\times10^{-14}\exp(27285/T)
\end{equation}
For the N$_2$-NH$_3$ system, the net reaction is: 
\begin{equation}
\mathrm{2NH_3 = N_2 + 3 H_2}
\end{equation}
and the chemical timescale is:
\begin{equation}
t_{\mathrm{NH_3}}=1.0\times10^{-7}p^{-1}\exp(52000/T)
\end{equation}
The chemical composition at the quenching level is fixed by the chemical equilibrium with the equilibrium constant\cite{zahnle_methane_2014}: 
\begin{equation}
K_{\mathrm{NH_3\cdot N_2}} = \frac{p\mathrm{NH_3}^2}{p\mathrm{N_2}\cdot(p\mathrm{H_2})^3}=5.90\times10^{-13}\exp(13207/T)
\end{equation}

Extended Figure A1 shows the temperature profiles of the 3D simulation at different latitudes and the quenching temperature profile (red dotted line) corresponding to $t_{mix}=t_{CO}$ from equation 3 for K$_{\mathrm{zz}}$ = 10$^8$ cm$^2$/s. The quenching levels correspond to the cross-over between the temperature profiles and the quenching temperature profile (large color dots). In addition, the mixing ratios of CH$_4$ and CO from chemical equilibrium (Equation 6) are indicated with solid black and gray lines, respectively.
Figure 1 shows instantaneous (no time average) mixing ratio maps of H$_2$O, CO, CH$_4$ and NH$_3$ above the quenching level. These maps were computed following the approach described above, computing the chemical composition at the quenching level for each point of the horizontal grid of the 3D model, and assuming uniform vertical mixing with K$_{\mathrm{zz}}$ = 10$^8$ cm$^2$/s. 
Extended Figure A2 shows latitudinal variations of the mixing ratios averaged longitudinally. The abundances of H$_2$O, CH$_4$ and NH$_3$ are reduced in the equatorial region (i.e. latitudes lower than 20$^\circ$). In contrast, the abundance of CO is enhanced in the equatorial region.

\subsection*{Rotational broadening}
We measured $v\mathrm{sin}i$ performing convolution of modeled emission spectra by the function of rotational broadening $G(\Delta\lambda)$ for a uniform star including limb darkening \cite{gray_observation_2005}:
\begin{equation}
    G(\Delta\lambda)=c_1 \left[1-\left( \Delta\lambda / \Delta\lambda_L\right)^2    \right]^{1/2} + c_2 \left[1-\left( \Delta\lambda / \Delta\lambda_L\right)^2    \right]
\end{equation}
with
\begin{equation}
c_1=\frac{2(1-\epsilon)}{\pi \Delta\lambda_L(1-\epsilon/3)}
\end{equation}
\begin{equation}
c_2=\frac{\epsilon}{2\Delta\lambda_L(1-\epsilon/3)}
\end{equation}
where $\epsilon$ is the limb darkening coefficient and $\Delta\lambda_L= \frac{\lambda_L}{c}v\mathrm{sin}i$ is the maximum Doppler shift for an equatorial velocity $v$ and an inclination $i$. Equation S10 is defined for $|{\Delta\lambda}|> \Delta\lambda_L$, so $G(|{\Delta\lambda}|> \Delta\lambda_L) =0$.
It can be generalized to a case with an equatorial depletion between latitudes $\pm \alpha$ (no emission contribution from low latitudes), assuming a high inclination ($\mathrm{sin}i \sim 1$). Following the calculation from \cite{gray_observation_2005} with the integration between the poles and latitudes $\pm \alpha$, the function of rotational broadening becomes:

\begin{multline}
    G_{partial}(\Delta\lambda, \alpha) = c_1 \left[ 1 - \left( \Delta \lambda / \Delta \lambda_L \right)^2   \right]^{1/2} + c_2 \left[ 1 - \left( \Delta \lambda / \Delta \lambda_L \right)^2    \right] \\
    + c_3 \left[ \frac{2(1-\epsilon)}{\epsilon} \mathrm{sin}(\alpha)+ \mathrm{sin}(\alpha)\left(1-\left( \Delta \lambda / \Delta \lambda_L \right)^2 - \mathrm{sin}(\alpha)^2\right)^{1/2} \right]\\
    +c_3 \left[1-\left( \Delta\lambda / \Delta \lambda_L \right)^2 \right] \mathrm{arcsin} \left( \frac{\mathrm{sin}(\alpha)}{\left[ 1 - \left( \Delta \lambda / \Delta \lambda_L \right)^2    \right]^{1/2}} \right) 
\end{multline}
where $c_3=-\frac{\epsilon }{\pi \Delta\lambda_L(1-\epsilon/3)}$

Figure 2B shows the function of rotational broadening normalised ($G_{partial}(\Delta\lambda=0, \alpha)=1$) for different values of $\alpha$. $G_{partial}(\Delta\lambda, \alpha=0)=G(\Delta\lambda)$, corresponding to a case with no equatorial depletion. For a depleted equatorial band with $\alpha>0$, the rotational broadening is reduced since there is no contribution from the equatorial region, associated to the maximum Doppler shift. The maximal radial velocity is equal to $\mathrm{cos}\alpha\times v\mathrm{sin}i$. As first approximation of the function of rotational broadening, we could propose $G_{partial}(\Delta\lambda, \alpha) \sim G(\Delta\lambda/\mathrm{cos}(\alpha))$ meaning that the measured rotational broadening with the $G$ function is $v\mathrm{sin}i \sim v\mathrm{sin}i_0\mathrm{cos}(\alpha)$, where $v\mathrm{sin}i_0$ is the maximum equatorial radial velocity. With this approximation, the rotational broadening is reduced by 6$\%$ for $\alpha=20^\circ$ compared to the case with no equatorial depletion.

However this approximation overestimates the rotational broadening. We found a better fit using 
$G_{partial}(\Delta\lambda, \alpha) \sim G\left(\Delta\lambda/\sqrt{(1-\mathrm{sin}(\alpha)^{3/2}}\right)$, meaning that the measured rotational broadening with the $G$ function is equal to: 
\begin{equation}
v\mathrm{sin}i \sim v\mathrm{sin}i_0 \sqrt{(1-\mathrm{sin}(\alpha)^{3/2})}
\end{equation}
This fit is illustrated on Figure 2B with dashed lines.
With this approximation, the rotational broadening is reduced by 10.6$\%$ for $\alpha=20^\circ$ compared to the case with no equatorial depletion. 

These calculations and this fit were performed assuming a high inclination with $\mathrm{sin}i \sim 1$, giving the equatorial band centered in the middle of the apparent disk. A first order correction can be added to take into account the inclination with:
\begin{equation}
v\mathrm{sin}i \sim v\mathrm{sin}i_0 \sqrt{(1-(\mathrm{sin}i\times\mathrm{sin}\alpha)^{3/2})}
\end{equation}
This corresponds to a geometric correction reducing the apparent extent of the equatorial band, and implying that the latitudinal extent of the equatorial depletion can be slightly underestimated with formula (14). However, the retrieved extent of the equatorial depletion is changed by less than 25$\%$ for $i>50^\circ$.

\subsection*{Validation with synthetic 3D spectra}
To test the detectability of latitudinal chemical variations, we computed idealized synthetic 3D spectra. We performed calculations using the petitRADTRANS radiative transfer code \cite{molliere_petitradtrans_2019} (also used in the Data analysis, see below). Spectra were 
computed at high resolution (R=10$^6$) in the K band for each point of a longitude-latitude grid (40$\times$40), taking the Doppler effect due to the rotation (with an equatorial velocity of 41 km/s and a null inclination) and the emission angle (interpolating among a linear grid of 20 values of cosine of the viewing angle). The total spectrum is the sum over the whole hemisphere of each spectrum, weighted by the cosine of the viewing angle, and convolved to a resolution of 10$^5$. For the atmospheric structure, we assumed a uniform vertical thermal profile, retrieved in the previous analysis of DENIS-J0255-470 
\citep{de_regt_eso_2024}. We considered an equatorial band between latitudes $\pm$20$^\circ$, in 
which the chemical composition differs from high latitudes. For high latitudes, we used the 
retrieved molecular abundances of \citep{de_regt_eso_2024} for H$_2$O, CO, CH$_4$ and NH$_3$, but 
by increasing the abundance of the latter two, respectively, by a factor of 5 and a factor of 3, to
match predictions from the 3D simulations (see Figure \ref{fig1}). We initially retrieved the 
apparent molecular $v\mathrm{sin}i$ for a full equatorial depletion of each species, taken 
separately. The retrieved $v\mathrm{sin}i$ are 36.4, 36.4, 36.2 and 36.4 km/s for, respectively, 
H$_2$O, CO, CH$_4$ and NH$_3$. These values are close to the expected $v\mathrm{sin}i$ of 36.67 km/
s from formula (1). We explored after the effect of a partial depletion or an enrichment, varying the abundance at low latitudes for each species separately, by a factor 0.1 to 10. The evolution of $v\mathrm{sin}i$ is shown in Figure A3.
Interestingly, we found that the apparent  $v\mathrm{sin}i$ varies more significantly for minor species (CH$_4$ and NH$_3$) than for major species (H$_2$O and CO). This can be explained by the fact that the molecular lines of minor species are shielded by the main absorbers. Thus, a change in their abundances has a much bigger impact on the spectra and their $v\mathrm{sin}i$. In contrast, H$_2$O is the main absorber at almost all wavelengths. An equatorial enrichment of H$_2$O has almost no effect on its 
$v\mathrm{sin}i$ because most of H$_2$O is already detectable. An equatorial depletion of H$_2$O by a factor greater than 5 is needed to reduce its $v\mathrm{sin}i$ by at least 1 km/s. Based on the predictions from the 3D simulation, only the $v\mathrm{sin}i$ of CH$_4$ and NH$_3$ should deviate significantly from the global $v\mathrm{sin}i$ of 41 km/s, with respectively 38.4 km/s and 39.5 km/s. However, given the enhanced sensitivity for minor species, and the fact the abundances for CH$_4$ and NH$_3$ in the synthetic 3D spectra are higher than those retrieved on DENIS J0255-4700, we could expect larger deviations of $v\mathrm{sin}i$ for CH$_4$ and NH$_3$.

\subsection*{Measurement of the rotation period from TESS data}
We measured the rotation period of DENIS J025503.3-470049 (DENIS J0255-4700) from TESS data. We made a cutout of TESS FFI time series for the region of the sky around DENIS J0255-4700 (https://mast.stsci.edu/tesscut/), with an area of 7$\times$4 pixels (see the top panel in Figure A7). DENIS J0255-4700 is faint in the TESS filter and a brighter background object is present at around 30 arcsec. The TESS data cover almost 55 days of continuous observation, providing enough accuracy to measure time variations of such a faint object. We performed a background substraction and we extracted a time serie, weighting pixels with a gaussian centered on the expected position of DENIS J0255-4700 and with $\sigma=1.2$. The periodogram (see the medium panel in Figure A7) indicates a peak at 2.21 hours, with a False Alarm Probability of $10^{-5}$. In addition, this period does not correspond to a peak in the TESS window function  (see the bottom panel in Figure A7). The detected peak is thus not due to TESS sampling, and we interpret it as the rotation period of DENIS J0255-4700. In contrast, peaks at 1.1 h, 1.5 h and 3.5 h are likely due to TESS sampling. A 2.21 h rotation period is consistent with a previous estimation of 1.7-2.2 hours \cite{koen_ic_2005}. It is also close but lower than the maximal rotation period of 2.32 hours derived from the measurements of $v\mathrm{sin}i$=41 km/s and radius R=0.78 R$_{Jup}$ \cite{de_regt_eso_2024}. Using these values, we estimate the inclination to be around 72$^{\circ}$, confirming a high inclination for DENIS J0255-4700.

\subsection*{Reduction of CRIRES data}
We used spectral observations of DENIS J0255-4700 that were taken, as part of the ESO SupJup Survey (Program ID: 110.23RW, PI: Snellen), with the CRyogenic high-resolution InfraRed Echelle Spectrograph (CRIRES) at the Very Large Telescope on the night of November 2, 2022. The chosen K-band wavelength-setting covers numerous absorption features from H$_2$O, CO, CH$_4$, and NH$_3$ and the employed $0.2^"$-slit provides a spectral resolution of $\mathcal{R}\sim$\,$10^5$. Twelve exposures of 300 seconds were carried out, summing up to a total integration time of 1 hour. The data were reduced with the excalibuhr\footnote{\url{https://excalibuhr.readthedocs.io/}} pipeline \cite{zhang_eso_2024}, following a standard procedure of dark-subtraction, flat-fielding, and removal of the sky background by subtracting AB nodding pairs from each other. An optimal extraction algorithm was applied to obtain the 1D spectra at each nodding position, which were subsequently mean-combined into one spectrum. The wavelength solution was refined with the help of telluric lines imprinted in the spectrum of a standard star, observed prior to DENIS J0255-4700. These same standard-star observations were used to divide out the telluric absorption and instrumental throughput from the DENIS J0255-4700 spectrum \cite{de_regt_eso_2024}. 
The poor correction in the deepest tellurics was mitigated by masking any pixels with transmission below $60\%$. In the end, the reduced and corrected spectrum yielded SNRs of $\sim$\,40 per pixel at the CO band ($2.34\ \mathrm{\mu m}$).

\subsection*{Analysis of CRIRES data}
We compared the observed spectrum of DENIS J0255-4700 to model spectra, adjusting the broadening to best fit the data. 
We started from the best fit of \cite{de_regt_eso_2024}, who carried out an atmospheric retrieval analysis and generated their model spectra (at R=$3\times10^5$) with the petitRADTRANS radiative transfer code \cite{molliere_petitradtrans_2019}. The line opacity of $^{12}$CO, $^{13}$CO \cite{li_rovibrational_2015}, H$_2$O \cite{polyansky_exomol_2018}, CH$_4$ \cite{hargreaves_accurate_2020}, NH$_3$ \cite{coles_exomol_2019}, CO$_2$ \cite{rothman_hitemp_2010}, and HCN \cite{harris_improved_2006} were included, but the separately fitted abundances did not constrain significant absorption from CO$_2$ and HCN. The best-fitting model found a global rotational velocity of $v\sin i=41.05^{+0.19}_{-0.19}\ \mathrm{km\ s^{-1}}$.
We used the model spectrum and the molecular contributions from this best fit but without broadening.
For the case with all molecules, we compared the observation spectrum to the model spectrum with all molecules.
For the cases with individual molecule, for instance CH$_4$, we extracted the observation residual: $F_{obs}-F_{model}^{no CH_4}$, where $F_{model}^{no CH_4}$ is broadened at $v\mathrm{sin}i=41.05$ km/s, and compared it to the model residual: $F_{model}^{full}-F_{model}^{no  CH_4}$ (no initial broadening for each model spectrum here). In this framework, we have to use a global rotational broadening to extract the observation residual.

To compute cross-correlation, all residual spectra were filtered to remove low frequencies, by subtracting spectra degraded at a spectral resolution of R=300, corresponding to a Gaussian convolution. The wavelengths corresponding to missing values were removed and we used templates with a larger spectral range than observations.
The cross-correlation (CCF) is defined as:
\begin{equation}
\mathrm{CCF}=\frac{\sum_{i} s_i t_{i,v}  }{\sqrt{ \left( \sum_{i} s_{i}^2 \right)  \left(\sum_{i} t_{i,0}^2 \right) } } 
\end{equation}
and the auto-correlation (ACF) is defined as:
\begin{equation}
\mathrm{ACF}=\frac{\sum_{i} t_{i,0} t_{i,v} }{\sum_{i} t_{i,0}^2 }
\end{equation}
where $i$ is the wavelength, $s$ the observation residual spectrum (or the observation spectrum when considering all molecules) and $t$ the model residual spectrum (or the model spectrum when considering all molecules) shifted at a given radial velocity $v$. The CCF and the ACF were computed in wavelength space.

To derive the signal-to-noise ratio of the cross-correlation, we used the following formula \cite{houlle_direct_2021}:
\begin{equation}
\mathrm{SNR}=\frac{\sum_{i} s_i t_{i,v} /\sigma_i ^2  }{\sqrt{ \sum_{i} t_{i,v}^2 /\sigma_i ^2} }
\end{equation}
where $\sigma^2$ the variance of the observed spectrum.

We also computed the log-likelihood using the following formula \cite{zucker_cross-correlation_2003}:
\begin{equation}
\log L = -\frac{N}{2}\log\left[1-\mathrm{C}^2 \right]
\end{equation}
Where $N$ is the number of spectral points and $\mathrm{C}=\mathrm{CCF}$ is the cross-correlation. 
This method is expected to work well for high-signal to noise spectra of isolated brown dwarfs and directly imaged planets \cite{brogi_retrieving_2019}. Figure 4 shows maps of delta log-likelihood ($\Delta\log L$ compared to the peak of log-likelihood) from the cross-correlation between observation and model residuals.

We estimated errors for the measurements of $v\mathrm{sin}i$ at the peak of the log-likelihood using \cite{zucker_cross-correlation_2003}:
\begin{equation}
\sigma_{v\mathrm{sin}i}^2=-\left[N \frac{\mathrm{C}''\mathrm{C}}{1-\mathrm{C}^2} \right]^{-1}
\end{equation}
where $\mathrm{C}$ is the cross-correlation at the peak of log-likelihood, $\mathrm{C}''$ is its second derivative (with respect to $v\mathrm{sin}i$), and N is the number of spectral points.
From this formula, we found errors (given in the main text and plotted in Figure 4) very similar to errors assuming that the log-likelihood was derived from a $\chi^2$, for which 1-$\sigma$ values correspond to $\Delta \log L$=-1/2.

For a more accurate evaluation of $v\mathrm{sin}i$ and to reduced systematics, we also perform calculations using the covariance matrix derived in \cite{de_regt_eso_2024}:
\begin{equation}
\Sigma_{0,ij} = \delta_{ij} + a^2 \sigma_{eff,ij}^2 \exp \left( -\frac{r_{ij}^2}{2l^2} \right)
\end{equation}
where $\delta_{ij}$ is the Kronecker delta, and $\sigma_i$ the flux-uncertainty of pixel $i$. Respectively, $a$ and $l$ are the correlation amplitude and length-scale. The pixel separation, $r_{ij}$, used in this study is given in the velocity space:
\begin{equation}
r_{ij} = 2 c \left|{\frac{\lambda_i-\lambda_j}{\lambda_i+\lambda_j}} \right|
\end{equation}
with c as the speed of light. The effective uncertainty, $\sigma_{eff,ij}$, used in this work is given by the arithmetic mean of the variances of pixels i and j:
\begin{equation}
\sigma_{eff,ij}^2=\frac{1}{2} \left(\sigma_{i}^2+\sigma_{j}^2\right)
\end{equation}
The correlation amplitude was derived for each order ($a_1=0.62$, $a_2=0.45$, $a_3=0.32$, $a_4=0.30$, $a_5=0.32$, $a_6=0.50$, $a_7=0.65$, with $l=23.87$ km/s).
Using the correlation matrix, the cross-correlation is expressed as:
\begin{equation}
\mathrm{CCF}=\frac{1}{s^2}M^T\Sigma_{0}^{-1}R 
\end{equation}
where $\Sigma_{0}^{-1}$ is the inverse of the covariance matrix, $M$ is the observation residual spectrum, $R$ is the shifted model residual and $s$ is a scaling factor defined (see \citep{de_regt_eso_2024}).
The top panel of Extended Figure A6 shows $v\mathrm{sin}i$ for each order, doing the calculation for each exposure and only plotting values with a SNR per exposure higher than 3. For NH$_3$, we fix the limit at a SNR combining all exposures higher than 5, in order not to eliminate all orders. The measurements correspond to the mean values from all exposures and the error bars come from the standard deviation from all exposures. 
According to this figure, H$_2$O is detected in all orders, CH$_4$ is detected in order 5, 6 and 7, while CO and NH$_3$ are only detected in orders 6 and 7. This motivated us to select only orders 5-7 for DMRB to limit the effects of systematics from other orders and to compare the rotational broadening of molecules in similar conditions.
H$_2$O (and so the full spectrum) shows little variability between orders, excepted in order 5. The derived values of CO shows little variability between orders 6 and 7. CH$_4$ shows variability between orders. However, its derived value combining orders 5-7 (2.228-2.472 $\mu$m) is mostly driven by the order 6 (2.321-2.369 $\mu$m) which has a SNR for CH$_4$ twice higher than orders 5 and 7.
The bottom panel of Extended Figure A6 shows $v\mathrm{sin}i$ combining measurements from order 5 to 7 and from each exposure, all taken separately. The measurements correspond to the mean values with the error bars coming from the standard deviation divided by the root square of the number of measurements.
This method tends to maximize the error bars compared to Figure 5 (based on formula (20)). The Extended Table 1 summarizes the results from the different methods, as well as measurements for different spectral resolutions of the low-frequency filtering (R=100, 300 and 1000). Although the values slightly change between the different analysis methods and resolutions of filtering, the conclusions are the same: a reduction of the $v\mathrm{sin}i$ for CH$_4$ and NH$_3$. The deviation for these molecules compared to the full model, H$_2$O or CO is always higher than 3 sigma.

\begin{table}
\caption{Measured $v\mathrm{sin}i$ for DENIS-J0255}
\label{tab1}
\begin{tabular}{@{}cccccccc@{}}
\toprule
 Orders & Exposures & Filtering &$v\mathrm{sin}i(\mathrm{all})$ & $v\mathrm{sin}i(\mathrm{CO})$ & $v\mathrm{sin}i(\mathrm{H_2O})$ & $v\mathrm{sin}i(\mathrm{CH_4})$ & $v\mathrm{sin}i(\mathrm{NH_3})$   \\ 
\midrule
 Combined & Combined & R=300 &$42.1_{-0.13}^{+0.13}$ & $41.2_{-0.19}^{+0.19}$ & $41.5_{-0.18}^{+0.18}$ & $37.0_{-0.40}^{+0.40}$ &  $31.8_{-1.48}^{+1.48}$\\ 
 Combined & Separated  & R=300 & $42.1_{-0.14}^{+0.14}$ & $41.3_{-0.18}^{+0.18}$ & $41.4_{-0.2}^{+0.2}$ & $37.2_{-0.55}^{+0.55}$ & $30.0_{-2.08}^{+2.08}$  \\ 
 Separated & Separated & R=300 & $41.7_{-0.22}^{+0.22}$ & $40.7_{-0.31}^{+0.31}$ & $40.8_{-0.27}^{+0.27}$ & $36.1_{-0.94}^{+0.94}$ & $27.5_{-1.9}^{+1.9}$  \\  
 Combined & Combined & R=100 &$42.1_{-0.13}^{+0.13}$ & $41.4_{-0.19}^{+0.19}$ & $41.5_{-0.19}^{+0.19}$ & $36.7_{-0.42}^{+0.42}$ &  $31.4_{-1.49}^{+1.49}$\\  
 Combined & Combined & R=1000 &$42.1_{-0.13}^{+0.13}$ & $41.1_{-0.19}^{+0.19}$ & $41.6_{-0.18}^{+0.18}$ & $37.2_{-0.41}^{+0.41}$ &  $31.9_{-1.47}^{+1.47}$\\  
 \hline
\end{tabular}
\end{table}


\subsection*{Applicability of DMRB to other targets}
We investigated the applicability of DMRB to other brown dwarfs. Similarly to the metric for Doppler Imaging \citep{crossfield_doppler_2014}, we propose a simple metric corresponding to a scaling of the detection of deviation of $v\mathrm{sin}i(\mathrm{CH4})$ from DENIS J0255-4700 (corresponding to a 10$\%$ deviation at $\sim$10 sigma) for observations in the K band. We express the metric as:
\begin{equation}
M= 10\times10^{(0.2\times(m_{K0}-m_K))}\times(D/D_0)\times \left(t/t_0\right)^{0.5}
\end{equation}
$M$ is expressed in sigma, $D$ is the telescope diameter, $D_0$=8 m, $t$ is the observing time, $t_0$= 1h.
This metric gives the detection level in sigma for a deviation of 10$\%$ of the $v\mathrm{sin}i$ of CH$_4$.
Figure A8 shows the cumulated number of accessible targets as a function of the apparent K magnitude (MJO) from the Ultracoolsheet \citep{best_ultracoolsheet_2025}. We limited the sample to objects with T$_{eff}$ $<$ 1500 K, close to the L-T transition for similar conditions as DENIS J0255-4700.
The vertical lines correspond to the apparent K magnitude to reach a 3-$\sigma$ detection (M=3) for a 8 m telescope (e.g. VLT) and a 39 m telescope (e.g. the ELT) for a 1-hour or a 5-hour observation. With a 8-m telescope, around 20 targets are accessible in 1 h while around 100 targets are accessible in 5 h. With the ELT, around 300 objects would be accessible for DMRB with just 1 h observations.

\backmatter

\bmhead{Supplementary information}
Not applicable

\bmhead{Acknowledgements}
We thank B. Bézard, C. Wilkinson, Hervé Bouy, T. Fouchet, T. Cavalié and T. Komacek for useful discussions. This work was granted access to the HPC resources of MesoPSL financed by the Region Ile de France and the project Equip@Meso (reference ANR-10-EQPX-29-01) of the programme Investissements d’Avenir supervised by the Agence Nationale pour la Recherche. This research has made use of the VizieR catalogue access tool, CDS, Strasbourg, France (DOI : 10.26093/cds/vizier). The original description  of the VizieR service was published in 2000, A$\&$AS 143, 23.

\section*{Declarations}

\begin{itemize}
\item Funding

This project has received funding from Agence Nationale de la Recherche (ANR) under grant ANR-23-CE31-0006-01 (MIRAGES).

\item Conflict of interest/Competing interests (check journal-specific guidelines for which heading to use)

Not applicable

\item Ethics approval and consent to participate

Not applicable

\item Consent for publication

Not applicable

\item Data availability 

The reduced spectrum of DENIS J0255-4700 from \cite{de_regt_eso_2024} and used in this study is available at the CDS via anonymous ftp to cdsarc.cds.unistra.fr (130.79.128.5) or via https://cdsarc.cds.unistra.fr/viz-bin/cat/J/A+A/688/A116

\item Materials availability

The reduced spectra and the best-fitting models used in this work are publicly available via Zenodo at https://doi.org/10.5281/zenodo.17112804.

\item Code availability 

Not applicable

\item Author contribution

B.~C. conceived the study and performed the data analysis. S.~de~R. performed the data reduction and provided the model spectra. M.~R., F.~K. and A.~D. contributed to the data analysis. L.~T. and B.~C. performed the 3D simulations and the calculation of the chemical composition maps.
B.~C. wrote the manuscript with inputs from all co-authors.

\end{itemize}

\begin{appendices}
\clearpage
\section{Extended figures}\label{secA1}

\begin{figure}[h] 
	\centering
	\includegraphics[width=0.8\textwidth]{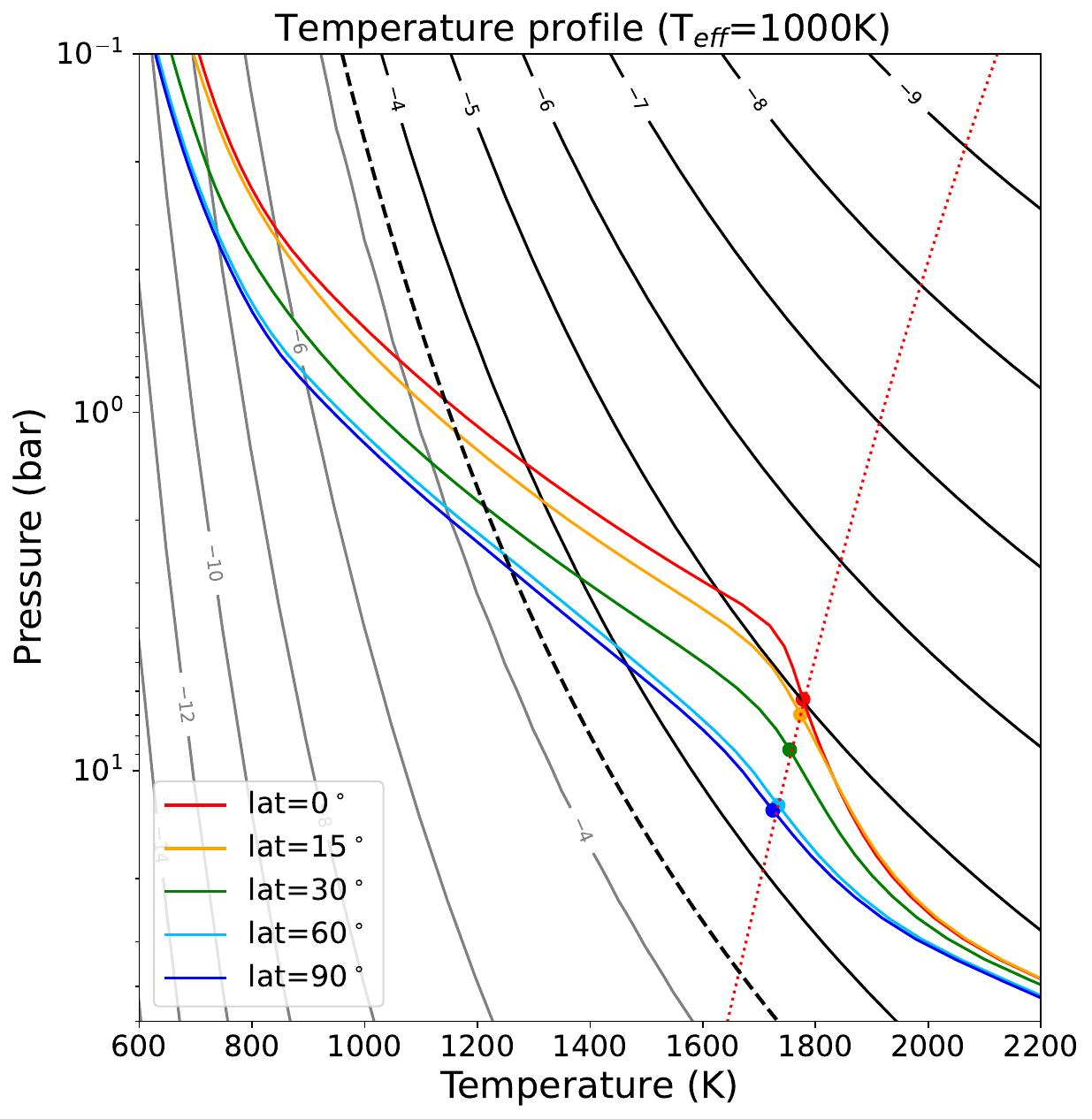}   
	\caption{
		Temperature profiles from the 3D simulation at different latitudes (color lines and profile of quenching temperature for the CH$_4$-CO conversion (red dashed line) assuming  K$_{\mathrm{zz}}$ = 10$^8$ cm$^2$/s.
        Quenching levels for the CH$_4$-CO conversion are indicated with large color dots. The mixing ratio of CH$_4$ (CO) from chemical equilibrium is shown with solid black (gray) lines. The temperature for which the abundances of CH$_4$ and CO are equal is indicated with a black dashed line.
        }
	\label{figE1} 
\end{figure}

\begin{figure}[h] 
	\centering
\includegraphics[width=0.8\textwidth]{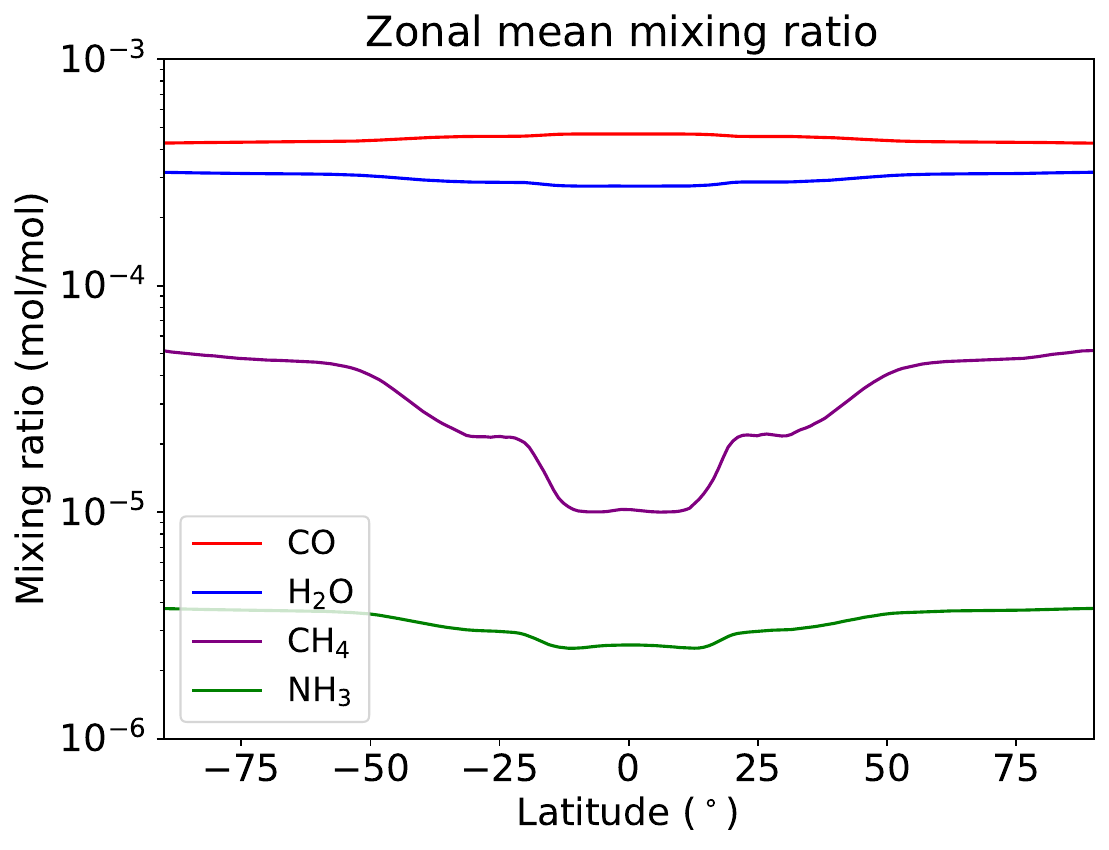}   
	\caption{Latitudinal variations of the mixing ratios of H$_2$O, CO, CH$_4$ and NH$_3$ from the 3D simulation and averaged longitudinally.}
	\label{figE2} 
\end{figure}

\begin{figure}[h] 
	\centering
\includegraphics[width=0.8\textwidth]{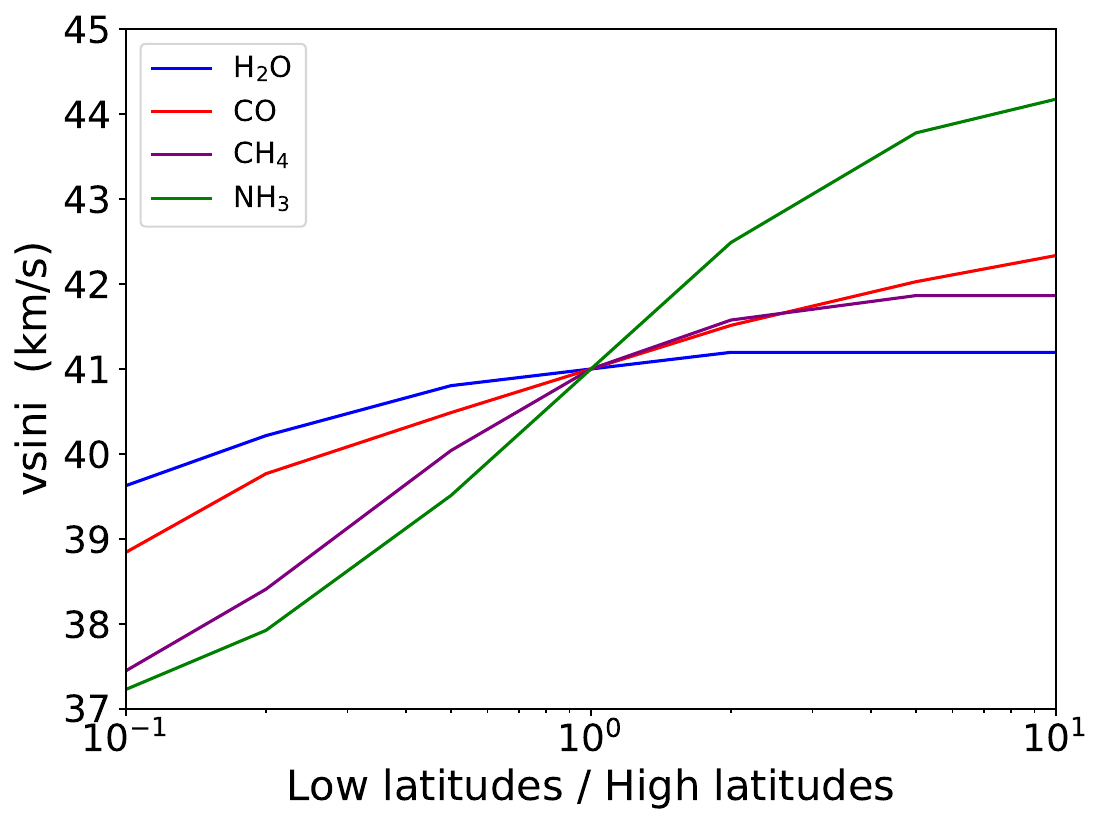}   
	\caption{Extracted $v\mathrm{sin}i$ of H$_2$O, CO, CH$_4$ and NH$_3$ from synthetic spectra computed with an equatorial band between latitudes $\pm$20$^\circ$. The x axis correspond to the ratio of molecular abundances at low latitudes by molecular abundances at high latitudes.}
	\label{figE3} 
\end{figure}

\begin{figure}[h] 
	\centering
\includegraphics[width=0.45\textwidth]{residual_CH4_R=300.pdf} 
\includegraphics[width=0.45\textwidth]{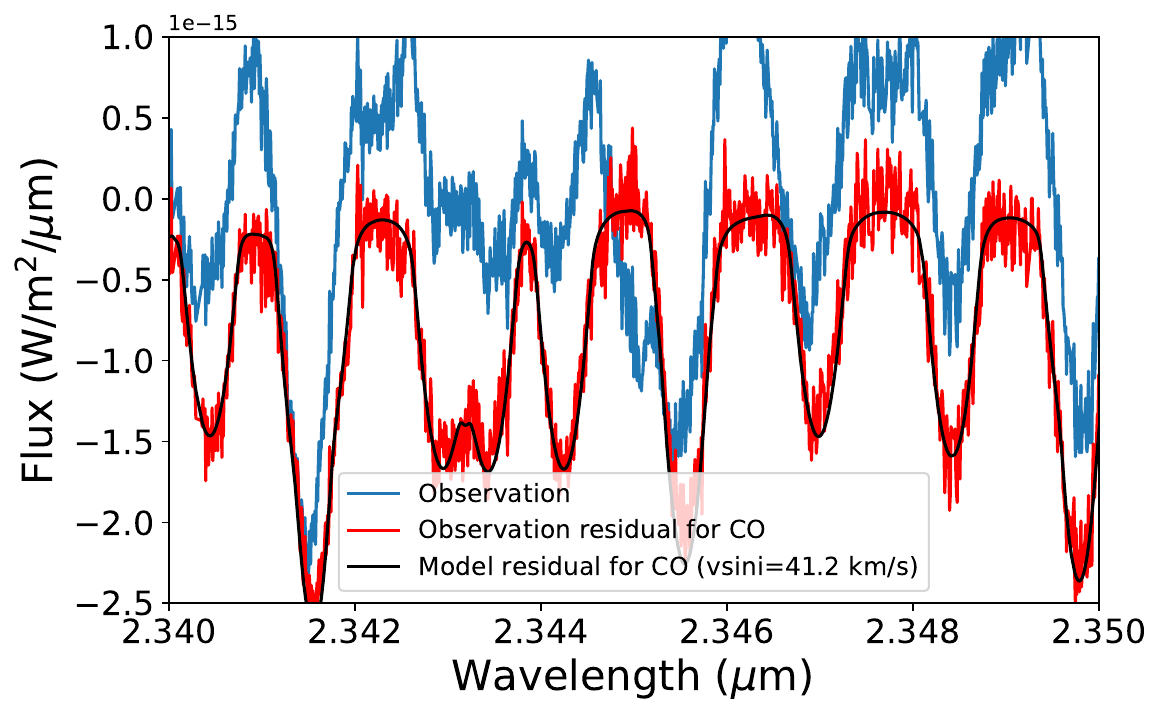} 
\includegraphics[width=0.45\textwidth]{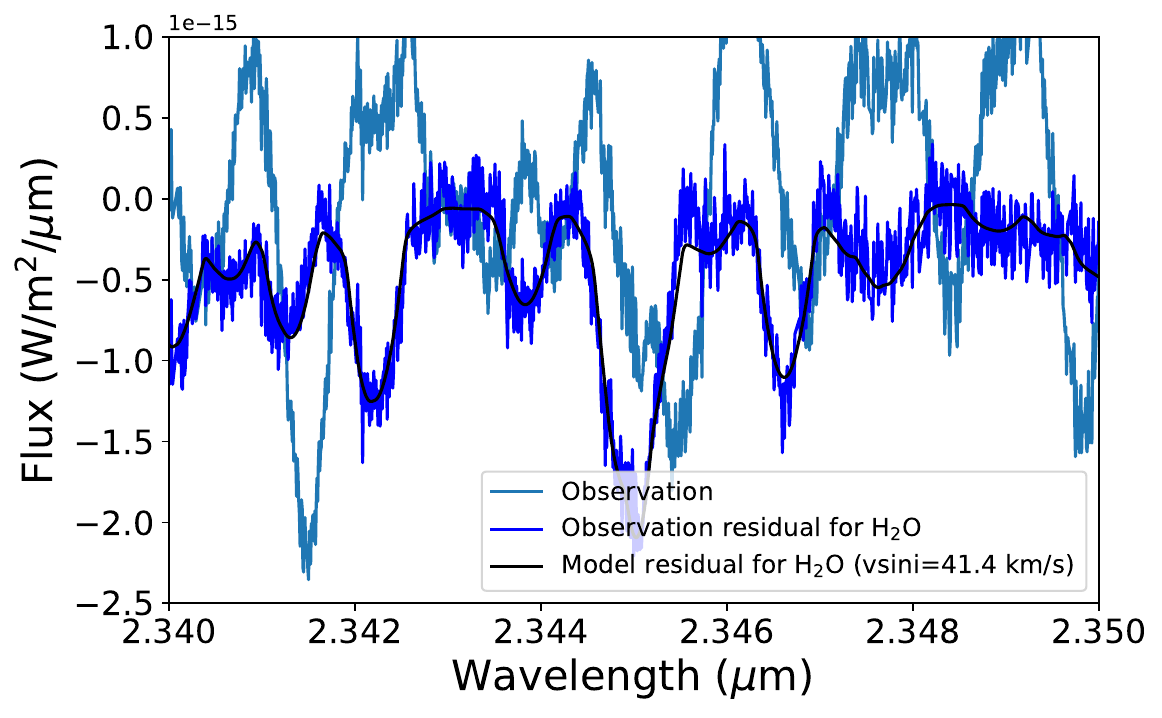} 
\includegraphics[width=0.45\textwidth]{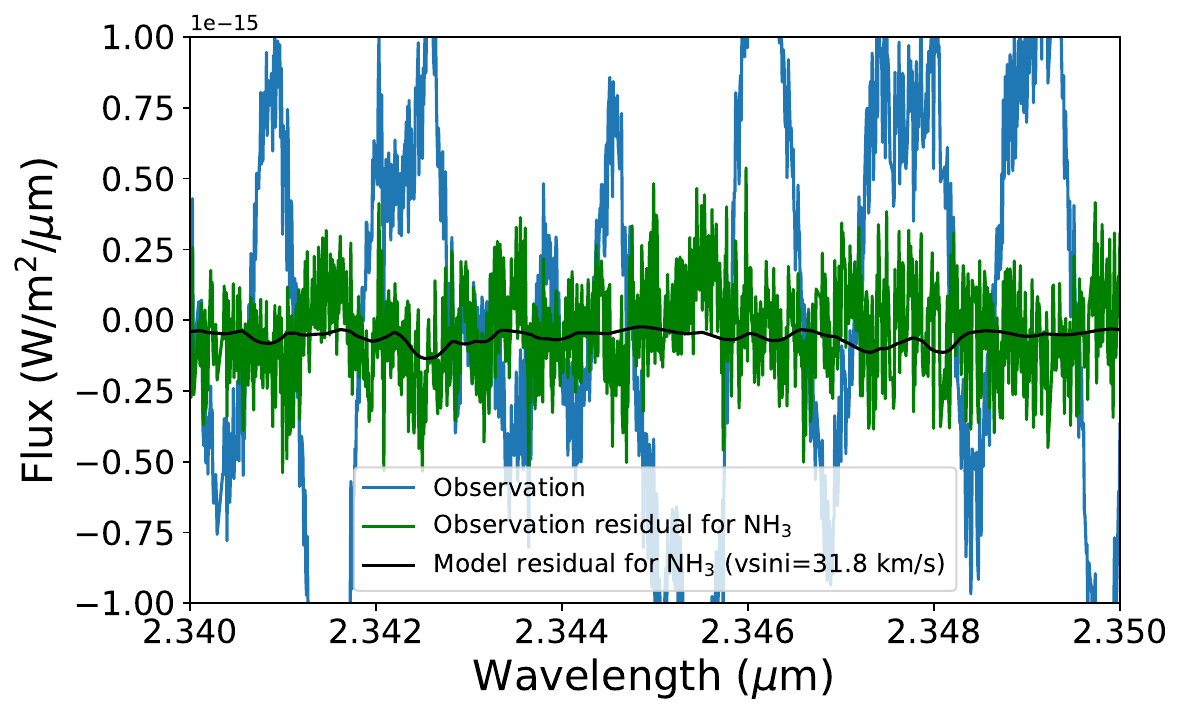} 
	\caption{
		Spectrum of DENIS J0255-4700 in the sixth spectral order of the CRIRES dataset after data reduction (light blue line). The comparison between the observation residual CH$_4$ (orange line, top left), CO (red line, top right), H$_2$O (blue line, top left) and CH$_4$ (green line, top left) and the model residual (green line, with the rotational broadening corresponding to the best fit) for each molecule is also indicated.    
        }
	\label{figE4} 
\end{figure}

\begin{figure}[h] 
	\centering
\includegraphics[width=0.8\textwidth]{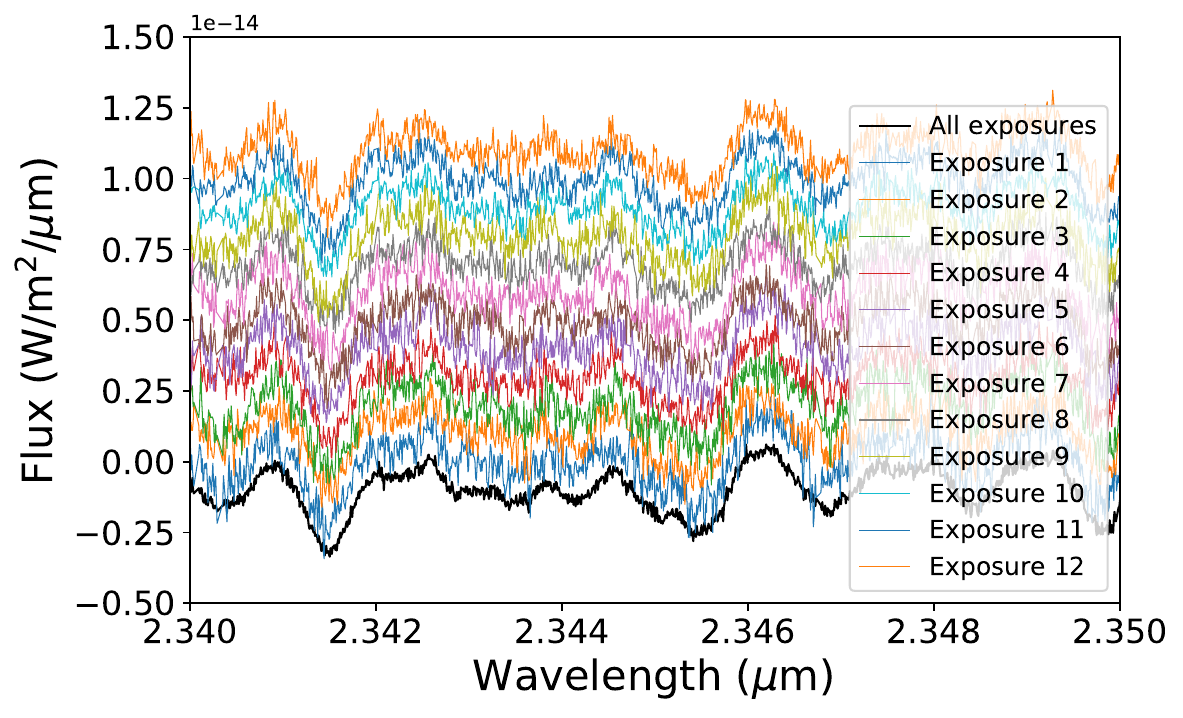}      
	\caption{Spectrum of DENIS J0255-4700 in the sixth spectral order of the CRIRES dataset after data reduction and after removing the low-frequencies (black line). The color lines represent the spectrum of each exposure, spaced at intervals of 10$^{-15}$ W/m$^2$/$\mu$m.}
	\label{figE5} 
\end{figure}

\begin{figure}[h] 
	\centering
\includegraphics[width=0.8\textwidth]{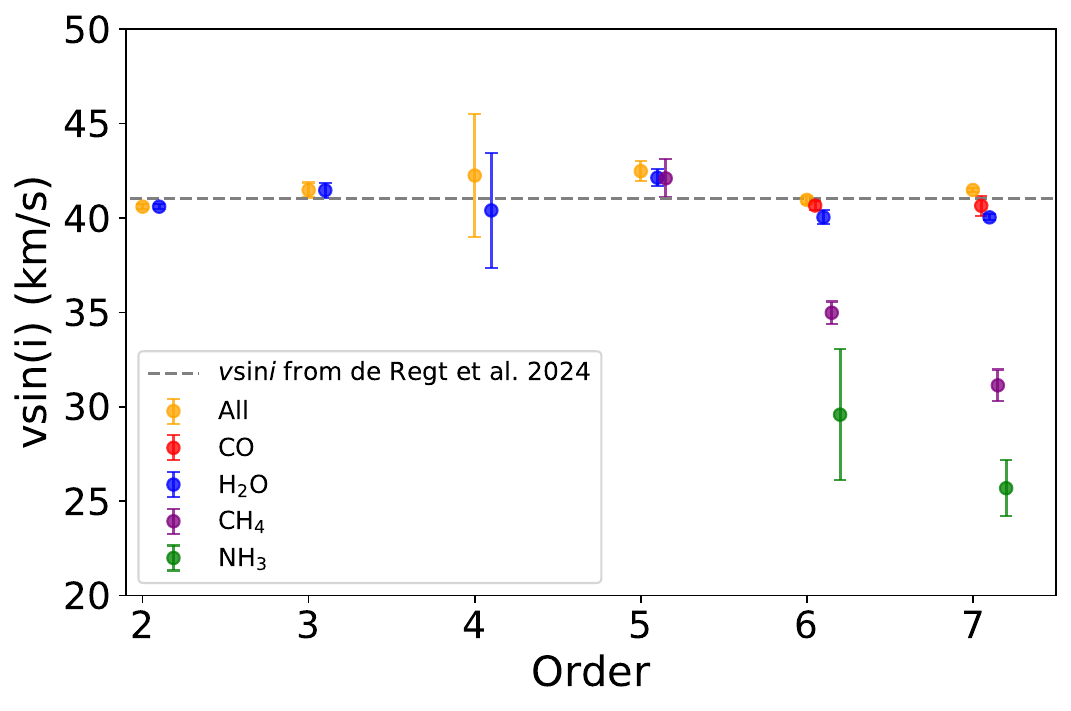}      
\includegraphics[width=0.8\textwidth]{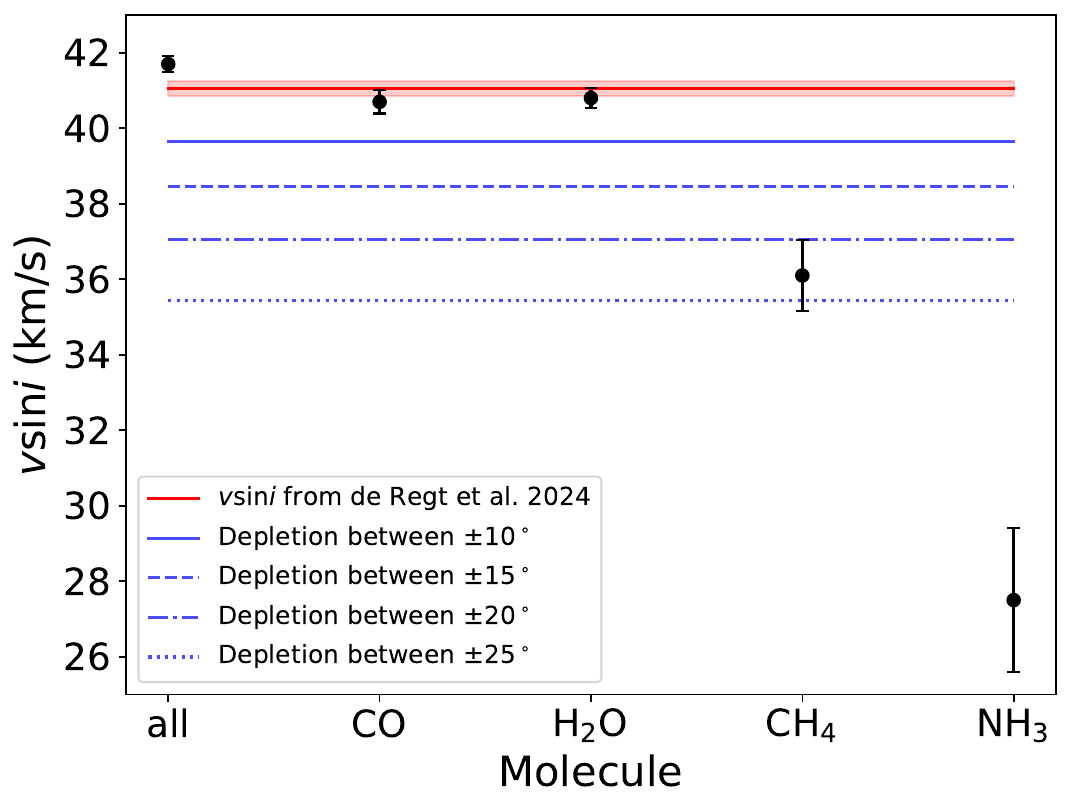}  
	\caption{Measurements of $v\mathrm{sin}i$ analyzing separately the 12 exposures and the orders of CRIRES data of DENIS J0255-4700. The top panel shows the measurements per molecule (colors) and per order (x axis). Only measurements with a SNR per exposure higher than 3 are indicated. The bottom panel is similar to Figure 5 with values corresponding to the mean between all exposures and  orders (considering only those with a significant detections, indicated in the top panel). The error bars where derived from the standard deviation between all exposures and orders. This tends to maximize error bars compared to Figure 5.
    }
	\label{figE6} 
\end{figure}

\begin{figure}[h] 
	\centering
\includegraphics[width=0.8\textwidth]{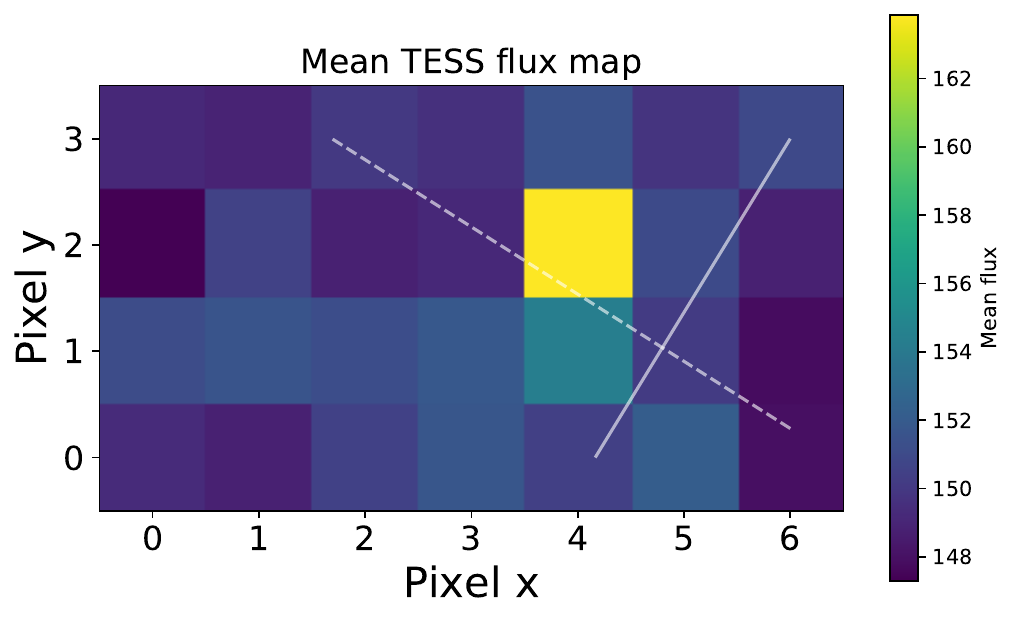}
\includegraphics[width=0.8\textwidth]{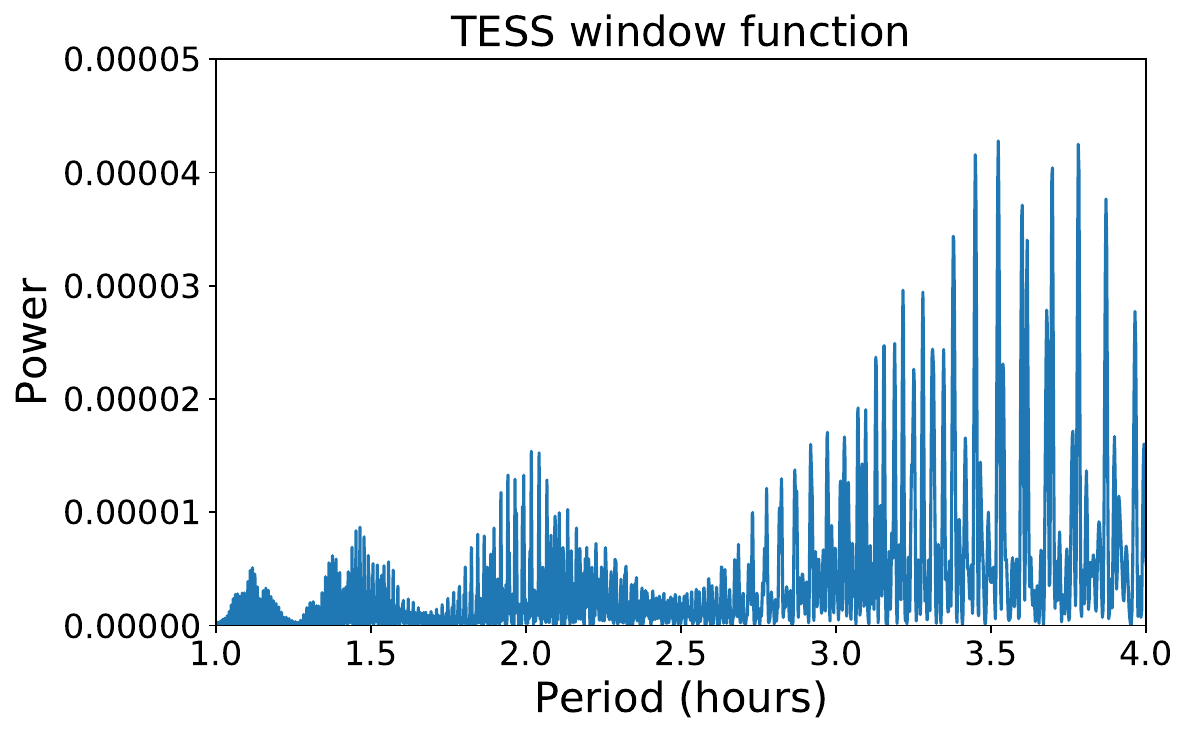}  
\includegraphics[width=0.8\textwidth]{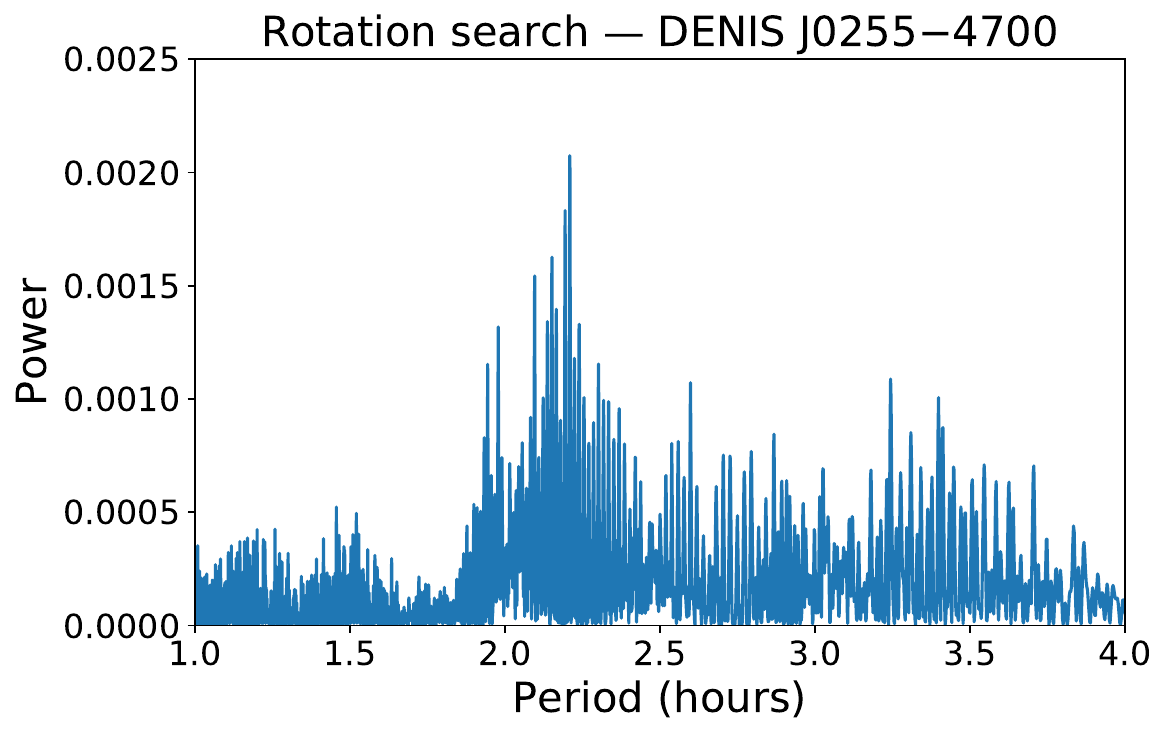}  
	\caption{TESS data for DENIS J0255-4700. 
    (top) Map of the mean flux from TESS in the sector of DENIS J0255-4700. The position of the brown dwarf is the crossing between the solid line and the dashed line (corresponding respectively to the right ascension and to the declination of DENIS J0255-4700). 
    (middel) Window function for TESS data from day 3933.66 to day 3988.28 (in BJD - 2457000).
    (bottom) Periodogram of TESS lightcurve for DENIS J0255-4700.
    }
	\label{figE7} 
\end{figure}

\begin{figure}[h] 
	\centering
\includegraphics[width=0.8\textwidth]{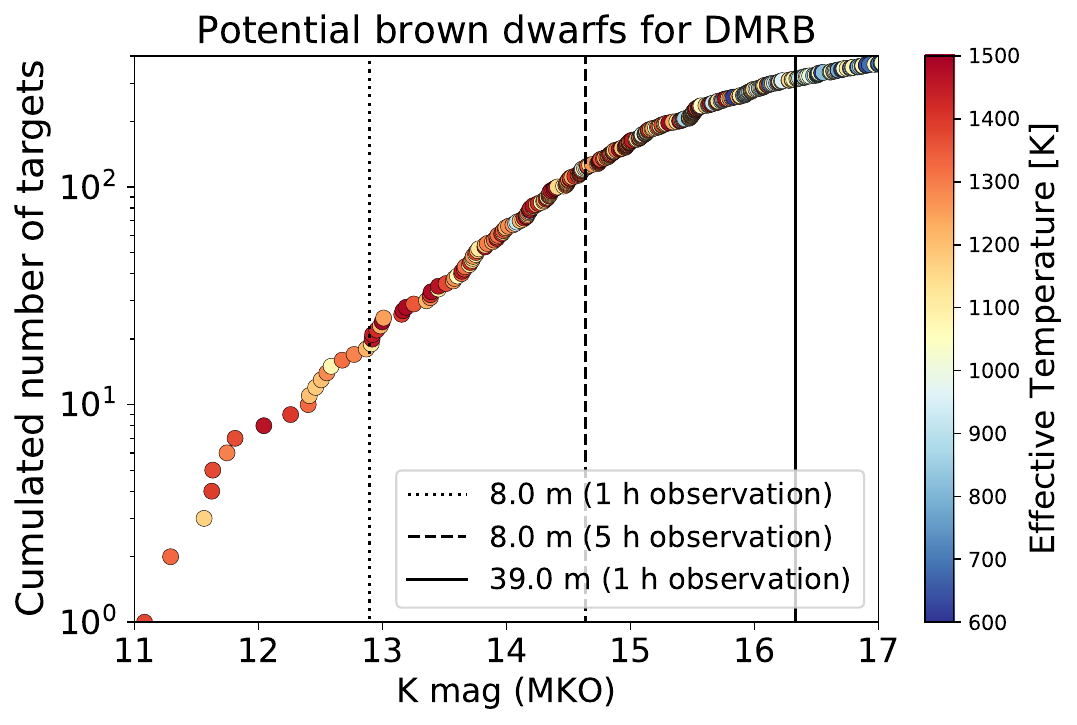}      
	\caption{Accessible brown dwarfs for DMRB. 
    Vertical lines correspond to the K magnitude limit for detectability at 3 sigma for 8 m or 39 m telescopes and for 1-h or 5-h observations. We used as sample the Ultracooldwarfsheet \cite{best_ultracoolsheet_2025}. The x-axis corresponds to the maximal K magnitude. The y-axis shows the cumulated number of accessible targets. The colors correspond to effective temperatures based on evolutionary models.}
	\label{figE8} 
\end{figure}

\end{appendices}

\clearpage

\bibliography{biblio_test}

@article{showman_atmospheric_2020,
	title = {Atmospheric {Dynamics} of {Hot} {Giant} {Planets} and {Brown} {Dwarfs}},
	volume = {216},
	issn = {0038-6308, 1572-9672},
	url = {http://link.springer.com/10.1007/s11214-020-00758-8},
	doi = {10.1007/s11214-020-00758-8},
	language = {en},
	number = {8},
	urldate = {2025-07-27},
	journal = {Space Science Reviews},
	author = {Showman, Adam P. and Tan, Xianyu and Parmentier, Vivien},
	month = dec,
	year = {2020},
	pages = {139},
}

@article{metchev_weather_2015,
	title = {{WEATHER} {ON} {OTHER} {WORLDS}. {II}. {SURVEY} {RESULTS}: {SPOTS} {ARE} {UBIQUITOUS} {ON} {L} {AND} {T} {DWARFS}},
	volume = {799},
	issn = {1538-4357},
	shorttitle = {{WEATHER} {ON} {OTHER} {WORLDS}. {II}. {SURVEY} {RESULTS}},
	url = {https://iopscience.iop.org/article/10.1088/0004-637X/799/2/154},
	doi = {10.1088/0004-637X/799/2/154},
	number = {2},
	urldate = {2025-07-27},
	journal = {The Astrophysical Journal},
	author = {Metchev, Stanimir A. and Heinze, Aren and Apai, Dániel and Flateau, Davin and Radigan, Jacqueline and Burgasser, Adam and Marley, Mark S. and Artigau, Étienne and Plavchan, Peter and Goldman, Bertrand},
	month = jan,
	year = {2015},
	pages = {154},
}

@article{tan_atmospheric_2021-1,
	title = {Atmospheric circulation of brown dwarfs and directly imaged exoplanets driven by cloud radiative feedback: effects of rotation},
	volume = {502},
	copyright = {http://creativecommons.org/licenses/by/4.0/},
	issn = {0035-8711, 1365-2966},
	shorttitle = {Atmospheric circulation of brown dwarfs and directly imaged exoplanets driven by cloud radiative feedback},
	url = {https://academic.oup.com/mnras/article/502/1/678/6089155},
	doi = {10.1093/mnras/stab060},
	language = {en},
	number = {1},
	urldate = {2025-07-27},
	journal = {Monthly Notices of the Royal Astronomical Society},
	author = {Tan, Xianyu and Showman, Adam P},
	month = jan,
	year = {2021},
	pages = {678--699},
}

@article{koen_ic_2005,
	title = {\textit{{I}}$_{\textrm{{C}}}$ and \textit{{R}}$_{\textrm{{C}}}$ band time-series observations of some bright ultracool dwarfs},
	volume = {360},
	issn = {00358711, 13652966},
	url = {https://academic.oup.com/mnras/article-lookup/doi/10.1111/j.1365-2966.2005.09119.x},
	doi = {10.1111/j.1365-2966.2005.09119.x},
	language = {en},
	number = {3},
	urldate = {2026-04-17},
	journal = {Monthly Notices of the Royal Astronomical Society},
	author = {Koen, Chris},
	month = jul,
	year = {2005},
	pages = {1132--1142},
}

@article{morales-calderon_sensitive_2006,
	title = {A {Sensitive} {Search} for {Variability} in {Late} {L} {Dwarfs}: {The} {Quest} for {Weather}},
	volume = {653},
	issn = {0004-637X},
	shorttitle = {A {Sensitive} {Search} for {Variability} in {Late} {L} {Dwarfs}},
	url = {https://iopscience.iop.org/article/10.1086/507866},
	doi = {10.1086/507866},
	language = {en},
	number = {2},
	urldate = {2026-04-17},
	journal = {The Astrophysical Journal},
	author = {Morales-Calderón, M. and Stauffer, J. R. and Kirkpatrick, J. Davy and Carey, S. and Gelino, C. R. and Navascués, D. Barrado Y and Rebull, L. and Lowrance, P. and Marley, M. S. and Charbonneau, D. and Patten, B. M. and Megeath, S. T. and Buzasi, D.},
	month = dec,
	year = {2006},
	pages = {1454},
}

@article{zhang_eso_2024,
	title = {The {ESO} {SupJup} {Survey}. {III}. {Confirmation} of$^{\textrm{13}}$ {CO} in {YSES} 1 b and {Atmospheric} {Detection} of {YSES} 1 c with {CRIRES}$^{\textrm{+}}$},
	volume = {168},
	issn = {0004-6256, 1538-3881},
	url = {https://iopscience.iop.org/article/10.3847/1538-3881/ad7ea9},
	doi = {10.3847/1538-3881/ad7ea9},
	language = {en},
	number = {6},
	urldate = {2026-04-17},
	journal = {The Astronomical Journal},
	author = {Zhang, Yapeng and González Picos, Darío and De Regt, Sam and Snellen, Ignas A. G. and Gandhi, Siddharth and Ginski, Christian and Kesseli, Aurora Y. and Landman, Rico and Mollière, Paul and Nasedkin, Evert and Sánchez-López, Alejandro and Stolker, Tomas and Inglis, Julie and Knutson, Heather A. and Mawet, Dimitri and Wallack, Nicole and Xuan, Jerry W.},
	month = dec,
	year = {2024},
	pages = {246},
}

@article{teinturier_clouds_2026,
	title = {Clouds as the driver of variability and colour changes in brown dwarf atmospheres},
	volume = {10},
	issn = {2397-3366},
	url = {https://www.nature.com/articles/s41550-025-02709-1},
	doi = {10.1038/s41550-025-02709-1},
	language = {en},
	number = {2},
	urldate = {2026-04-17},
	journal = {Nature Astronomy},
	author = {Teinturier, Lucas and Charnay, Benjamin and Spiga, Aymeric and Bézard, Bruno},
	month = jan,
	year = {2026},
	pages = {224--233},
}

@article{rodriguez-ovalle_stratospheric_2024,
	title = {Stratospheric aerosols and {C}$_{\textrm{6}}$ {H}$_{\textrm{6}}$ in {Jupiter}’s south polar region from {JWST}/{MIRI} observations},
	volume = {691},
	copyright = {https://creativecommons.org/licenses/by/4.0},
	issn = {0004-6361, 1432-0746},
	url = {https://www.aanda.org/10.1051/0004-6361/202451453},
	doi = {10.1051/0004-6361/202451453},
	language = {en},
	urldate = {2026-04-23},
	journal = {Astronomy \& Astrophysics},
	author = {Rodríguez-Ovalle, Pablo and Guerlet, Sandrine and Fouchet, Thierry and Harkett, Jake and Cavalié, Thibault and Hue, Vincent and Vinatier, Sandrine and López-Puertas, Manuel and Fletcher, Leigh N. and Lellouch, Emmanuel and Hueso, Ricardo and De Pater, Imke and Orton, Glenn S. and Roman, Michael T. and Hammel, Heidi B. and Milam, Stefanie N. and King, Oliver R. T.},
	month = nov,
	year = {2024},
	pages = {A51},
}

@article{rodriguez-ovalle_jwst_2025,
	title = {{JWST} observations of exogenic species on {Jupiter}: {HCN}, {H}$_{\textrm{2}}$ {O}, and {CO}$_{\textrm{2}}$},
	volume = {696},
	copyright = {https://creativecommons.org/licenses/by/4.0},
	issn = {0004-6361, 1432-0746},
	shorttitle = {{JWST} observations of exogenic species on {Jupiter}},
	url = {https://www.aanda.org/10.1051/0004-6361/202453575},
	doi = {10.1051/0004-6361/202453575},
	language = {en},
	urldate = {2026-04-23},
	journal = {Astronomy \& Astrophysics},
	author = {Rodríguez-Ovalle, Pablo and Fouchet, Thierry and Cavalié, Thibault and Lellouch, Emmanuel and Fletcher, Leigh N. and Harkett, Jake and Hue, Vincent and Benmahi, Bilal and De Pater, Imke},
	month = apr,
	year = {2025},
	pages = {A173},
}

@article{tan_large-amplitude_2025,
	title = {Large-amplitude variability driven by giant dust storms on a planetary-mass companion},
	language = {en},
	journal = {Science Advances},
	author = {Tan, Xianyu and Zhang, Xi and Marley, Mark S and Zhou, Yifan and Lew, Ben W P and Miles, Brittany E and Batalha, Natasha E and Biller, Beth A and Chauvin, Gaël and Hinkley, Sasha and Hoch, Kielan K W and Manjavacas, Elena and Metchev, Stanimir and Petrus, Simon and Rickman, Emily and Skemer, Andrew and Suárez, Genaro and Sutlieff, Ben J and Vos, Johanna M and Whiteford, Niall},
	year = {2025},
}

@article{crossfield_doppler_2014,
	title = {Doppler imaging of exoplanets and brown dwarfs},
	volume = {566},
	issn = {0004-6361, 1432-0746},
	url = {http://www.aanda.org/10.1051/0004-6361/201423750},
	doi = {10.1051/0004-6361/201423750},
	language = {en},
	urldate = {2026-07-07},
	journal = {Astronomy \& Astrophysics},
	author = {Crossfield, Ian J. M.},
	month = jun,
	year = {2014},
	pages = {A130},
}

@misc{best_ultracoolsheet_2025,
	title = {The {UltracoolSheet}: {Photometry}, {Astrometry}, {Spectroscopy}, and {Multiplicity} for 4000+ {Ultracool} {Dwarfs} and {Imaged} {Exoplanets}},
	copyright = {Creative Commons Attribution 4.0 International},
	shorttitle = {The {UltracoolSheet}},
	url = {https://zenodo.org/doi/10.5281/zenodo.4169084},
	doi = {10.5281/ZENODO.4169084},
	language = {en},
	urldate = {2026-07-07},
	publisher = {Zenodo},
	author = {Best, William M. J. and Dupuy, Trent J. and Liu, Michael C. and Sanghi, Aniket and Siverd, Robert J. and Zhang, Zhoujian},
	month = jul,
	year = {2025},
}

@article{zahnle_methane_2014,
	title = {Methane, {Carbon} {Monoxide}, and {Ammonia} in {Brown} {Dwarfs} and {Self}-{Luminous} {Giant} {Planets}},
	volume = {797},
	doi = {10.1088/0004-637X/797/1/41},
	journal = {The Astrophysical Journal},
	author = {Zahnle, K. J. and Marley, M. S.},
	month = dec,
	year = {2014},
	note = {\_eprint: 1408.6283},
	pages = {41},
}

@article{fortney_beyond_2020,
	title = {Beyond {Equilibrium} {Temperature}: {How} the {Atmosphere}/{Interior} {Connection} {Affects} the {Onset} of {Methane}, {Ammonia}, and {Clouds} in {Warm} {Transiting} {Giant} {Planets}},
	volume = {160},
	doi = {10.3847/1538-3881/abc5bd},
	number = {6},
	journal = {The Astronomical Journal},
	author = {Fortney, Jonathan J. and Visscher, Channon and Marley, Mark S. and Hood, Callie E. and Line, Michael R. and Thorngren, Daniel P. and Freedman, Richard S. and Lupu, Roxana},
	month = dec,
	year = {2020},
	note = {\_eprint: 2010.00146},
	pages = {288},
}

@article{suarez_ultracool_2023,
	title = {Ultracool {Dwarfs} {Observed} with the {Spitzer} {Infrared} {Spectrograph}: {Equatorial} {Latitudes} in {L} {Dwarf} {Atmospheres} {Are} {Cloudier}},
	volume = {954},
	issn = {2041-8205, 2041-8213},
	shorttitle = {Ultracool {Dwarfs} {Observed} with the {Spitzer} {Infrared} {Spectrograph}},
	url = {https://iopscience.iop.org/article/10.3847/2041-8213/acec4b},
	doi = {10.3847/2041-8213/acec4b},
	number = {1},
	urldate = {2025-07-27},
	journal = {The Astrophysical Journal Letters},
	author = {Suárez, Genaro and Vos, Johanna M. and Metchev, Stanimir and Faherty, Jacqueline K. and Cruz, Kelle},
	month = sep,
	year = {2023},
	pages = {L6},
}

@article{vos_spitzer_2020,
	title = {Spitzer {Variability} {Properties} of {Low}-gravity {L} {Dwarfs}},
	volume = {160},
	issn = {0004-6256, 1538-3881},
	url = {https://iopscience.iop.org/article/10.3847/1538-3881/ab9642},
	doi = {10.3847/1538-3881/ab9642},
	number = {1},
	urldate = {2025-07-27},
	journal = {The Astronomical Journal},
	author = {Vos, Johanna M. and Biller, Beth A. and Allers, Katelyn N. and Faherty, Jacqueline K. and Liu, Michael C. and Metchev, Stanimir and Eriksson, Simon and Manjavacas, Elena and Dupuy, Trent J. and Janson, Markus and Radigan-Hoffman, Jacqueline and Crossfield, Ian and Bonnefoy, Mickaël and Best, William M. J. and Homeier, Derek and Schlieder, Joshua E. and Brandner, Wolfgang and Henning, Thomas and Bonavita, Mariangela and Buenzli, Esther},
	month = jul,
	year = {2020},
	pages = {38},
}

@article{vos_viewing_2017,
	title = {The {Viewing} {Geometry} of {Brown} {Dwarfs} {Influences} {Their} {Observed} {Colors} and {Variability} {Amplitudes}},
	volume = {842},
	issn = {0004-637X, 1538-4357},
	url = {https://iopscience.iop.org/article/10.3847/1538-4357/aa73cf},
	doi = {10.3847/1538-4357/aa73cf},
	number = {2},
	urldate = {2025-07-27},
	journal = {The Astrophysical Journal},
	author = {Vos, Johanna M. and Allers, Katelyn N. and Biller, Beth A.},
	month = jun,
	year = {2017},
	pages = {78},
}

@article{lee_dynamically_2024,
	title = {Dynamically coupled kinetic chemistry in brown dwarf atmospheres – {II}. {Cloud} and chemistry connections in directly imaged sub-{Jupiter} exoplanets},
	volume = {529},
	copyright = {https://creativecommons.org/licenses/by/4.0/},
	issn = {0035-8711, 1365-2966},
	url = {https://academic.oup.com/mnras/article/529/3/2686/7611715},
	doi = {10.1093/mnras/stae537},
	language = {en},
	number = {3},
	urldate = {2025-07-27},
	journal = {Monthly Notices of the Royal Astronomical Society},
	author = {Lee, Elspeth K H and Tan, Xianyu and Tsai, Shang-Min},
	month = mar,
	year = {2024},
	pages = {2686--2701},
}

@article{morley_spectral_2014,
	title = {Spectral {Variability} from the {Patchy} {Atmospheres} of {T} and {Y} {Dwarfs}},
	volume = {789},
	doi = {10.1088/2041-8205/789/1/L14},
	journal = {The Astrophysical Journal Letters},
	author = {Morley, C. V. and Marley, M. S. and Fortney, J. J. and Lupu, R.},
	month = jul,
	year = {2014},
	note = {\_eprint: 1406.0863},
	pages = {L14},
}

@article{charnay_self-consistent_2018,
	title = {A {Self}-consistent {Cloud} {Model} for {Brown} {Dwarfs} and {Young} {Giant} {Exoplanets}: {Comparison} with {Photometric} and {Spectroscopic} {Observations}},
	volume = {854},
	doi = {10.3847/1538-4357/aaac7d},
	number = {2},
	journal = {Astrophys. J.},
	author = {Charnay, B. and Bézard, B. and Baudino, J. -L. and Bonnefoy, M. and Boccaletti, A. and Galicher, R.},
	month = feb,
	year = {2018},
	note = {\_eprint: 1711.11483},
	pages = {172},
}

@book{gray_observation_2005,
	edition = {3},
	title = {The {Observation} and {Analysis} of {Stellar} {Photospheres}},
	copyright = {https://www.cambridge.org/core/terms},
	isbn = {978-0-521-85186-2 978-0-521-06681-5 978-1-316-03657-0},
	url = {https://www.cambridge.org/core/product/identifier/9781316036570/type/book},
	doi = {10.1017/CBO9781316036570},
	urldate = {2025-07-27},
	publisher = {Cambridge University Press},
	author = {Gray, David F.},
	month = nov,
	year = {2005},
}

@article{mohanty_rotation_2003,
	title = {Rotation and {Activity} in {Mid}-{M} to {L} {Field} {Dwarfs}},
	volume = {583},
	issn = {0004-637X, 1538-4357},
	url = {https://iopscience.iop.org/article/10.1086/345097},
	doi = {10.1086/345097},
	language = {en},
	number = {1},
	urldate = {2026-04-17},
	journal = {The Astrophysical Journal},
	author = {Mohanty, Subhanjoy and Basri, Gibor},
	month = jan,
	year = {2003},
	pages = {451--472},
}

@article{zapatero_osorio_spectroscopic_2006,
	title = {Spectroscopic {Rotational} {Velocities} of {Brown} {Dwarfs}},
	volume = {647},
	issn = {0004-637X, 1538-4357},
	url = {https://iopscience.iop.org/article/10.1086/505484},
	doi = {10.1086/505484},
	language = {en},
	number = {2},
	urldate = {2026-04-17},
	journal = {The Astrophysical Journal},
	author = {Zapatero Osorio, M. R. and Martin, E. L. and Bouy, H. and Tata, R. and Deshpande, R. and Wainscoat, R. J.},
	month = aug,
	year = {2006},
	pages = {1405--1412},
}

@article{de_regt_eso_2024,
	title = {The {ESO} {SupJup} {Survey}: {I}. {Chemical} and isotopic characterisation of the late {L}-dwarf {DENIS} {J0255}-4700 with {CRIRES}$^{\textrm{+}}$},
	volume = {688},
	copyright = {https://creativecommons.org/licenses/by/4.0},
	issn = {0004-6361, 1432-0746},
	shorttitle = {The {ESO} {SupJup} {Survey}},
	url = {https://www.aanda.org/10.1051/0004-6361/202348508},
	doi = {10.1051/0004-6361/202348508},
	urldate = {2025-07-27},
	journal = {Astronomy \& Astrophysics},
	author = {De Regt, S. and Gandhi, S. and Snellen, I. A. G. and Zhang, Y. and Ginski, C. and González Picos, D. and Kesseli, A. Y. and Landman, R. and Mollière, P. and Nasedkin, E. and Sánchez-López, A. and Stolker, T.},
	month = aug,
	year = {2024},
	pages = {A116},
}

@article{palle_ground-breaking_2025,
	title = {Ground-breaking exoplanet science with the {ANDES} spectrograph at the {ELT}},
	volume = {59},
	issn = {0922-6435, 1572-9508},
	url = {https://link.springer.com/10.1007/s10686-025-10000-4},
	doi = {10.1007/s10686-025-10000-4},
	language = {en},
	number = {3},
	urldate = {2025-07-27},
	journal = {Experimental Astronomy},
	author = {Palle, Enric and Biazzo, Katia and Bolmont, Emeline and Mollière, Paul and Poppenhaeger, Katja and Birkby, Jayne and Brogi, Matteo and Chauvin, Gael and Chiavassa, Andrea and Hoeijmakers, Jens and Lellouch, Emmanuel and Lovis, Christophe and Maiolino, Roberto and Nortmann, Lisa and Parviainen, Hannu and Pino, Lorenzo and Turbet, Martin and Weder, Jesse and Albrecht, Simon and Antoniucci, Simone and Barros, Susana C. and Beaudoin, Andre and Benneke, Bjorn and Boisse, Isabelle and Bonomo, Aldo S. and Borsa, Francesco and Brandeker, Alexis and Brandner, Wolfgang and Buchhave, Lars A. and Cheffot, Anne-Laure and Deborde, Robin and Debras, Florian and Doyon, Rene and Di Marcantonio, Paolo and Giacobbe, Paolo and González Hernández, Jonay I. and Helled, Ravit and Kreidberg, Laura and Machado, Pedro and Maldonado, Jesus and Marconi, Alessandro and Martins, B. L. Canto and Miceli, Adriano and Mordasini, Christoph and N’Diaye, Mamadou and Niedzielski, Andrzej and Nisini, Brunella and Origlia, Livia and Peroux, Celine and Pietrow, Alexander G. M. and Pinna, Enrico and Rauscher, Emily and Reffert, Sabine and Rodríguez-López, Cristina and Rousselot, Philippe and Sanna, Nicoletta and Santos, Nuno C. and Simonnin, Adrien and Suárez Mascareño, Alejandro and Zanutta, Alessio and Zapatero-Osorio, Maria Rosa and Zechmeister, Mathias},
	month = jun,
	year = {2025},
	pages = {29},
}

@article{kasper_pcs_2021,
	title = {{PCS} — {A} {Roadmap} for {Exoearth} {Imaging} with the {ELT}},
	volume = {pp. 38-43},
	copyright = {Copyright European Southern Observatory},
	issn = {0722-6691},
	url = {http://doi.eso.org/10.18727/0722-6691/5221},
	doi = {10.18727/0722-6691/5221},
	urldate = {2025-07-27},
	journal = {Published in The Messenger vol. 182},
	publisher = {European Southern Observatory (ESO)},
	author = {Kasper, Markus and Cerpa Urra, Nelly and Pathak, Prashant and Bonse, Markus and Nousiainen, Jalo and Engler, Byron and Heritier, Cédric Taïssir and Kammerer, Jens and Leveratto, Serban and Rajani, Chang and Bristow, Paul and Le Louarn, Miska and Madec, Pierre-Yves and Ströbele, Stefan and Verinaud, Christophe and Glauser, Adrian and Quanz, Sascha P. and Helin, Tapio and Keller, Christoph and Snik, Frans and Boccaletti, Anthony and Chauvin, Gaël and Mouillet, David and Kulcsár, Caroline and Raynaud, Henri-François},
	year = {2021},
	note = {Artwork Size: 6 pages
Medium: PDF},
	pages = {6 pages},
}

@inproceedings{brandl_instrument_2010,
	address = {San Diego, California, USA},
	title = {Instrument concept and science case for the mid-{IR} {E}-{ELT} imager and spectrograph {METIS}},
	url = {http://proceedings.spiedigitallibrary.org/proceeding.aspx?doi=10.1117/12.857346},
	doi = {10.1117/12.857346},
	urldate = {2025-07-27},
	author = {Brandl, Bernhard R. and Lenzen, Rainer and Pantin, Eric and Glasse, Alistair and Blommaert, Joris and Venema, Lars and Molster, Frank and Siebenmorgen, Ralf and Kendrew, Sarah and Baes, Maarten and Böhnhardt, Hermann and Brandner, Wolfgang and Van Dishoeck, Ewine and Henning, Thomas and Käufl, Hans Ullrich and Lagage, Pierre-Olivier and Moore, Toby J. T. and Waelkens, Christoffel and Van Der Werf, Paul},
	editor = {McLean, Ian S. and Ramsay, Suzanne K. and Takami, Hideki},
	month = jul,
	year = {2010},
	pages = {77352G},
}

@article{sanghavi_photopolarimetric_2018,
	title = {Photopolarimetric {Characteristics} of {Brown} {Dwarfs}. {I}. {Uniform} {Cloud} {Decks}},
	volume = {866},
	issn = {0004-637X, 1538-4357},
	url = {https://iopscience.iop.org/article/10.3847/1538-4357/aadf94},
	doi = {10.3847/1538-4357/aadf94},
	number = {1},
	urldate = {2025-07-27},
	journal = {The Astrophysical Journal},
	author = {Sanghavi, Suniti and Shporer, Avi},
	month = oct,
	year = {2018},
	pages = {28},
}

@article{molliere_petitradtrans_2019,
	title = {{petitRADTRANS}: {A} {Python} radiative transfer package for exoplanet characterization and retrieval},
	volume = {627},
	copyright = {https://www.edpsciences.org/en/authors/copyright-and-licensing},
	issn = {0004-6361, 1432-0746},
	shorttitle = {{petitRADTRANS}},
	url = {https://www.aanda.org/10.1051/0004-6361/201935470},
	doi = {10.1051/0004-6361/201935470},
	language = {en},
	urldate = {2025-09-08},
	journal = {Astronomy \& Astrophysics},
	author = {Mollière, P. and Wardenier, J. P. and Van Boekel, R. and Henning, Th. and Molaverdikhani, K. and Snellen, I. A. G.},
	month = jul,
	year = {2019},
	pages = {A67},
}

@article{blain_spectralmodel_2024,
	title = {{SpectralModel}: a high-resolution framework {forpetitRADTRANS} 3},
	volume = {9},
	copyright = {http://creativecommons.org/licenses/by/4.0/},
	issn = {2475-9066},
	shorttitle = {{SpectralModel}},
	url = {https://joss.theoj.org/papers/10.21105/joss.07028},
	doi = {10.21105/joss.07028},
	language = {en},
	number = {102},
	urldate = {2025-09-08},
	journal = {Journal of Open Source Software},
	author = {Blain, Doriann and Mollière, Paul and Nasedkin, Evert},
	month = oct,
	year = {2024},
	pages = {7028},
}

@article{baraffe_evolutionary_2003,
	title = {Evolutionary models for cool brown dwarfs and extrasolar giant planets. {The} case of {HD} 209458},
	volume = {402},
	doi = {10.1051/0004-6361:20030252},
	journal = {Astronomy \& Astrophysics},
	author = {Baraffe, I. and Chabrier, G. and Barman, T. S. and Allard, F. and Hauschildt, P. H.},
	month = may,
	year = {2003},
	note = {\_eprint: astro-ph/0302293},
	pages = {701--712},
}

@article{zucker_cross-correlation_2003,
	title = {Cross-correlation and maximum-likelihood analysis: a new approach to combining cross-correlation functions},
	volume = {342},
	issn = {0035-8711, 1365-2966},
	shorttitle = {Cross-correlation and maximum-likelihood analysis},
	url = {https://academic.oup.com/mnras/article/342/4/1291/957971},
	doi = {10.1046/j.1365-8711.2003.06633.x},
	language = {en},
	number = {4},
	urldate = {2025-09-04},
	journal = {Monthly Notices of the Royal Astronomical Society},
	author = {Zucker, S.},
	month = jul,
	year = {2003},
	pages = {1291--1298},
}

@article{allers_measurement_2020,
	title = {A measurement of the wind speed on a brown dwarf},
	volume = {368},
	issn = {0036-8075, 1095-9203},
	url = {https://www.science.org/doi/10.1126/science.aaz2856},
	doi = {10.1126/science.aaz2856},
	language = {en},
	number = {6487},
	urldate = {2025-07-27},
	journal = {Science},
	author = {Allers, Katelyn. N. and Vos, Johanna M. and Biller, Beth A. and Williams, Peter. K. G.},
	month = apr,
	year = {2020},
	pages = {169--172},
}

@article{faherty_methane_2024,
	title = {Methane emission from a cool brown dwarf},
	volume = {628},
	issn = {0028-0836, 1476-4687},
	url = {https://www.nature.com/articles/s41586-024-07190-w},
	doi = {10.1038/s41586-024-07190-w},
	language = {en},
	number = {8008},
	urldate = {2025-07-27},
	journal = {Nature},
	author = {Faherty, Jacqueline K. and Burningham, Ben and Gagné, Jonathan and Suárez, Genaro and Vos, Johanna M. and Alejandro Merchan, Sherelyn and Morley, Caroline V. and Rowland, Melanie and Lacy, Brianna and Kiman, Rocio and Caselden, Dan and Kirkpatrick, J. Davy and Meisner, Aaron and Schneider, Adam C. and Kuchner, Marc Jason and Bardalez Gagliuffi, Daniella Carolina and Beichman, Charles and Eisenhardt, Peter and Gelino, Christopher R. and Gharib-Nezhad, Ehsan and Gonzales, Eileen and Marocco, Federico and Rothermich, Austin James and Whiteford, Niall},
	month = apr,
	year = {2024},
	pages = {511--514},
}

@article{chen_global_2024,
	title = {Global weather map reveals persistent top-of-atmosphere features on the nearest brown dwarfs},
	volume = {533},
	copyright = {https://creativecommons.org/licenses/by/4.0/},
	issn = {0035-8711, 1365-2966},
	url = {https://academic.oup.com/mnras/article/533/3/3114/7737682},
	doi = {10.1093/mnras/stae1995},
	language = {en},
	number = {3},
	urldate = {2025-07-27},
	journal = {Monthly Notices of the Royal Astronomical Society},
	author = {Chen, Xueqing and Biller, Beth A and Vos, Johanna M and Crossfield, Ian J M and Mace, Gregory N and Hood, Callie E and Tan, Xianyu and Allers, Katelyn N and Martin, Emily C and Bubb, Emma and Fortney, Jonathan J and Morley, Caroline V and Hammond, Mark},
	month = aug,
	year = {2024},
	pages = {3114--3143},
}

@article{vogt_doppler_1987,
	title = {Doppler images of rotating stars using maximum entropy image reconstruction},
	volume = {321},
	issn = {0004-637X, 1538-4357},
	url = {http://adsabs.harvard.edu/doi/10.1086/165647},
	doi = {10.1086/165647},
	language = {en},
	urldate = {2025-07-27},
	journal = {The Astrophysical Journal},
	author = {Vogt, Steven S. and Penrod, G. Donald and Hatzes, Artie P.},
	month = oct,
	year = {1987},
	pages = {496},
}

@article{hargreaves_accurate_2020,
	title = {An {Accurate}, {Extensive}, and {Practical} {Line} {List} of {Methane} for the {HITEMP} {Database}},
	volume = {247},
	issn = {0067-0049, 1538-4365},
	url = {https://iopscience.iop.org/article/10.3847/1538-4365/ab7a1a},
	doi = {10.3847/1538-4365/ab7a1a},
	language = {en},
	number = {2},
	urldate = {2025-09-16},
	journal = {The Astrophysical Journal Supplement Series},
	author = {Hargreaves, Robert J. and Gordon, Iouli E. and Rey, Michael and Nikitin, Andrei V. and Tyuterev, Vladimir G. and Kochanov, Roman V. and Rothman, Laurence S.},
	month = apr,
	year = {2020},
	pages = {55},
}

@article{li_rovibrational_2015,
	title = {{ROVIBRATIONAL} {LINE} {LISTS} {FOR} {NINE} {ISOTOPOLOGUES} {OF} {THE} {CO} {MOLECULE} {IN} {THE} \textit{{X}}$^{\textrm{1}}$ sigma$^{\textrm{+}}$ {GROUND} {ELECTRONIC} {STATE}},
	volume = {216},
	copyright = {http://iopscience.iop.org/info/page/text-and-data-mining},
	issn = {1538-4365},
	url = {https://iopscience.iop.org/article/10.1088/0067-0049/216/1/15},
	doi = {10.1088/0067-0049/216/1/15},
	language = {en},
	number = {1},
	urldate = {2025-09-16},
	journal = {The Astrophysical Journal Supplement Series},
	author = {Li, Gang and Gordon, Iouli E. and Rothman, Laurence S. and Tan, Yan and Hu, Shui-Ming and Kassi, Samir and Campargue, Alain and Medvedev, Emile S.},
	month = jan,
	year = {2015},
	pages = {15},
}

@article{harris_improved_2006,
	title = {Improved {HCN}/{HNC} linelist, model atmospheres and synthetic spectra for {WZ} {Cas}},
	volume = {367},
	issn = {0035-8711, 1365-2966},
	url = {https://academic.oup.com/mnras/article-lookup/doi/10.1111/j.1365-2966.2005.09960.x},
	doi = {10.1111/j.1365-2966.2005.09960.x},
	language = {en},
	number = {1},
	urldate = {2025-09-16},
	journal = {Monthly Notices of the Royal Astronomical Society},
	author = {Harris, G. J. and Tennyson, J. and Kaminsky, B. M. and Pavlenko, Y. V. and Jones, H. R. A.},
	month = mar,
	year = {2006},
	pages = {400--406},
}

@article{polyansky_exomol_2018,
	title = {{ExoMol} molecular line lists {XXX}: a complete high-accuracy line list for water},
	volume = {480},
	copyright = {http://creativecommons.org/licenses/by/4.0/},
	issn = {0035-8711, 1365-2966},
	shorttitle = {{ExoMol} molecular line lists {XXX}},
	url = {https://academic.oup.com/mnras/article/480/2/2597/5054049},
	doi = {10.1093/mnras/sty1877},
	language = {en},
	number = {2},
	urldate = {2025-09-16},
	journal = {Monthly Notices of the Royal Astronomical Society},
	author = {Polyansky, Oleg L and Kyuberis, Aleksandra A and Zobov, Nikolai F and Tennyson, Jonathan and Yurchenko, Sergei N and Lodi, Lorenzo},
	month = oct,
	year = {2018},
	pages = {2597--2608},
}

@article{rothman_hitemp_2010,
	title = {{HITEMP}, the high-temperature molecular spectroscopic database},
	volume = {111},
	doi = {10.1016/j.jqsrt.2010.05.001},
	journal = {JQSRT},
	author = {Rothman, L. S. and Gordon, I. E. and Barber, R. J. and Dothe, H. and Gamache, R. R. and Goldman, A. and Perevalov, V. I. and Tashkun, S. A. and Tennyson, J.},
	month = oct,
	year = {2010},
	pages = {2139--2150},
}

@article{houlle_direct_2021,
	title = {Direct imaging and spectroscopy of exoplanets with the {ELT}/{HARMONI} high-contrast module},
	volume = {652},
	copyright = {https://creativecommons.org/licenses/by/4.0},
	issn = {0004-6361, 1432-0746},
	url = {https://www.aanda.org/10.1051/0004-6361/202140479},
	doi = {10.1051/0004-6361/202140479},
	language = {en},
	urldate = {2025-09-04},
	journal = {Astronomy \& Astrophysics},
	author = {Houllé, M. and Vigan, A. and Carlotti, A. and Choquet, É. and Cantalloube, F. and Phillips, M. W. and Sauvage, J.-F. and Schwartz, N. and Otten, G. P. P. L. and Baraffe, I. and Emsenhuber, A. and Mordasini, C.},
	month = aug,
	year = {2021},
	pages = {A67},
}

@article{brogi_retrieving_2019,
	title = {Retrieving {Temperatures} and {Abundances} of {Exoplanet} {Atmospheres} with {High}-{Resolution} {Cross}-{Correlation} {Spectroscopy}},
	volume = {157},
	issn = {0004-6256, 1538-3881},
	url = {http://arxiv.org/abs/1811.01681},
	doi = {10.3847/1538-3881/aaffd3},
	language = {en},
	number = {3},
	urldate = {2025-09-04},
	journal = {The Astronomical Journal},
	author = {Brogi, Matteo and Line, Michael R.},
	month = mar,
	year = {2019},
	note = {arXiv:1811.01681 [astro-ph]},
	pages = {114},
}

@article{kirkpatrick_new_2005,
	title = {New {Spectral} {Types} {L} and {T}},
	volume = {43},
	doi = {10.1146/annurev.astro.42.053102.134017},
	journal = {Annual Review of Astronomy \& Astrophysics},
	author = {Kirkpatrick, J. D.},
	month = sep,
	year = {2005},
	pages = {195--245},
}

@article{cushing_atmospheric_2008,
	title = {Atmospheric {Parameters} of {Field} {L} and {T} {Dwarfs1}},
	volume = {678},
	issn = {0004-637X, 1538-4357},
	url = {https://iopscience.iop.org/article/10.1086/526489},
	doi = {10.1086/526489},
	language = {en},
	number = {2},
	urldate = {2025-07-27},
	journal = {The Astrophysical Journal},
	author = {Cushing, Michael C. and Marley, Mark S. and Saumon, D. and Kelly, Brandon C. and Vacca, William D. and Rayner, John T. and Freedman, Richard S. and Lodders, Katharina and Roellig, Thomas L.},
	month = may,
	year = {2008},
	pages = {1372--1395},
}

@article{marley_patchy_2010,
	title = {A {Patchy} {Cloud} {Model} for the {L} to {T} {Dwarf} {Transition}},
	volume = {723},
	doi = {10.1088/2041-8205/723/1/L117},
	journal = {The Astrophysical Journal Letters},
	author = {Marley, M. S. and Saumon, D. and Goldblatt, C.},
	month = nov,
	year = {2010},
	note = {\_eprint: 1009.6217},
	pages = {L117--L121},
}

@article{showman_atmospheric_2013,
	title = {{ATMOSPHERIC} {DYNAMICS} {OF} {BROWN} {DWARFS} {AND} {DIRECTLY} {IMAGED} {GIANT} {PLANETS}},
	volume = {776},
	copyright = {http://iopscience.iop.org/info/page/text-and-data-mining},
	issn = {0004-637X, 1538-4357},
	url = {https://iopscience.iop.org/article/10.1088/0004-637X/776/2/85},
	doi = {10.1088/0004-637X/776/2/85},
	number = {2},
	urldate = {2025-07-27},
	journal = {The Astrophysical Journal},
	author = {Showman, Adam P. and Kaspi, Yohai},
	month = oct,
	year = {2013},
	pages = {85},
}

@article{zhang_atmospheric_2014,
	title = {{ATMOSPHERIC} {CIRCULATION} {OF} {BROWN} {DWARFS}: {JETS}, {VORTICES}, {AND} {TIME} {VARIABILITY}},
	volume = {788},
	copyright = {http://iopscience.iop.org/info/page/text-and-data-mining},
	issn = {2041-8205, 2041-8213},
	shorttitle = {{ATMOSPHERIC} {CIRCULATION} {OF} {BROWN} {DWARFS}},
	url = {https://iopscience.iop.org/article/10.1088/2041-8205/788/1/L6},
	doi = {10.1088/2041-8205/788/1/L6},
	number = {1},
	urldate = {2025-07-27},
	journal = {The Astrophysical Journal},
	author = {Zhang, Xi and Showman, Adam P.},
	month = may,
	year = {2014},
	pages = {L6},
}

@article{showman_atmospheric_2019,
	title = {Atmospheric {Circulation} of {Brown} {Dwarfs} and {Jupiter}- and {Saturn}-like {Planets}: {Zonal} {Jets}, {Long}-term {Variability}, and {QBO}-type {Oscillations}},
	volume = {883},
	issn = {0004-637X, 1538-4357},
	shorttitle = {Atmospheric {Circulation} of {Brown} {Dwarfs} and {Jupiter}- and {Saturn}-like {Planets}},
	url = {https://iopscience.iop.org/article/10.3847/1538-4357/ab384a},
	doi = {10.3847/1538-4357/ab384a},
	number = {1},
	urldate = {2025-07-27},
	journal = {The Astrophysical Journal},
	author = {Showman, Adam P. and Tan, Xianyu and Zhang, Xi},
	month = sep,
	year = {2019},
	pages = {4},
}

@article{tan_atmospheric_2021,
	title = {Atmospheric circulation of brown dwarfs and directly imaged exoplanets driven by cloud radiative feedback: global and equatorial dynamics},
	volume = {502},
	copyright = {http://creativecommons.org/licenses/by/4.0/},
	issn = {0035-8711, 1365-2966},
	shorttitle = {Atmospheric circulation of brown dwarfs and directly imaged exoplanets driven by cloud radiative feedback},
	url = {https://academic.oup.com/mnras/article/502/2/2198/6101225},
	doi = {10.1093/mnras/stab097},
	language = {en},
	number = {2},
	urldate = {2025-07-27},
	journal = {Monthly Notices of the Royal Astronomical Society},
	author = {Tan, Xianyu and Showman, Adam P},
	month = feb,
	year = {2021},
	pages = {2198--2219},
}

@article{apai_zones_2017,
	title = {Zones, spots, and planetary-scale waves beating in brown dwarf atmospheres},
	volume = {357},
	doi = {10.1126/science.aam9848},
	journal = {Science},
	author = {Apai, D. and Karalidi, T. and Marley, M. S. and Yang, H. and Flateau, D. and Metchev, S. and Cowan, N. B. and Buenzli, E. and Burgasser, A. J. and Radigan, J. and Artigau, E. and Lowrance, P.},
	month = aug,
	year = {2017},
	pages = {683--687},
}

@article{biller_time_2017,
	title = {The time domain for brown dwarfs and directly imaged giant exoplanets: the power of variability monitoring},
	volume = {13},
	doi = {10.1080/21672857.2017.1303105},
	journal = {The Astronomical Review},
	author = {Biller, B.},
	month = jan,
	year = {2017},
	pages = {1--27},
}

@article{crossfield_global_2014,
	title = {A global cloud map of the nearest known brown dwarf},
	volume = {505},
	doi = {10.1038/nature12955},
	journal = {Nature},
	author = {Crossfield, I. J. M. and Biller, B. and Schlieder, J. E. and Deacon, N. R. and Bonnefoy, M. and Homeier, D. and Allard, F. and Buenzli, E. and Henning, T. and Brandner, W. and Goldman, B. and Kopytova, T.},
	month = jan,
	year = {2014},
	note = {\_eprint: 1401.8145},
	pages = {654--656},
}

@article{wordsworth_gliese_2011,
	title = {Gliese 581d is the {First} {Discovered} {Terrestrial}-mass {Exoplanet} in the {Habitable} {Zone}},
	volume = {733},
	journal = {The Astrophysical Journal Letters},
	author = {Wordsworth, R. D. and Forget, F. and Selsis, F. and Millour, E. and Charnay, B. and Madeleine, J.-B.},
	month = jun,
	year = {2011},
	pages = {L48},
}

@article{teinturier_radiative_2024,
	title = {The radiative and dynamical impact of clouds in the atmosphere of the hot {Jupiter} {WASP}-43 b},
	volume = {683},
	copyright = {https://creativecommons.org/licenses/by/4.0},
	issn = {0004-6361, 1432-0746},
	url = {https://www.aanda.org/10.1051/0004-6361/202347069},
	doi = {10.1051/0004-6361/202347069},
	language = {en},
	urldate = {2024-10-01},
	journal = {Astronomy \& Astrophysics},
	author = {Teinturier, L. and Charnay, B. and Spiga, A. and Bézard, B. and Leconte, J. and Mechineau, A. and Ducrot, E. and Millour, E. and Clément, N.},
	month = mar,
	year = {2024},
	pages = {A231},
}

@article{coles_exomol_2019,
	title = {{ExoMol} molecular line lists - {XXXV}. {A} rotation-vibration line list for hot ammonia},
	volume = {490},
	doi = {10.1093/mnras/stz2778},
	number = {4},
	journal = {Monthly Notices of the Royal Astronomical Society},
	author = {Coles, Phillip A. and Yurchenko, Sergei N. and Tennyson, Jonathan},
	month = dec,
	year = {2019},
	note = {\_eprint: 1911.10369},
	pages = {4638--4647},
}

@article{spiga_global_2020,
	title = {Global climate modeling of {Saturn}'s atmosphere. {Part} {II}: {Multi}-annual high-resolution dynamical simulations},
	volume = {335},
	issn = {00191035},
	shorttitle = {Global climate modeling of {Saturn}'s atmosphere. {Part} {II}},
	url = {https://linkinghub.elsevier.com/retrieve/pii/S0019103518306912},
	doi = {10.1016/j.icarus.2019.07.011},
	language = {en},
	urldate = {2025-09-02},
	journal = {Icarus},
	author = {Spiga, Aymeric and Guerlet, Sandrine and Millour, Ehouarn and Indurain, Mikel and Meurdesoif, Yann and Cabanes, Simon and Dubos, Thomas and Leconte, Jérémy and Boissinot, Alexandre and Lebonnois, Sébastien and Sylvestre, Mélody and Fouchet, Thierry},
	month = jan,
	year = {2020},
	pages = {113377},
}

\end{document}